\documentclass[preprint]{aastex}
\usepackage{graphicx}
\usepackage{amssymb}
\usepackage{amsmath}
\usepackage{color}
\usepackage{ulem}

\def\ie{i.e.,~}

\begin{document}

\title{Vorticity, Shocks and Magnetic Fields in Subsonic, ICM-like Turbulence}
\author{David H. Porter\altaffilmark{1}, T. W. Jones\altaffilmark{1,2}, and Dongsu Ryu\altaffilmark{3}}
\altaffiltext{1}{Minnesota Supercomputing Institute, University of Minnesota, Minneapolis, MN 5545; dhp@umn.edu, twj@umn.edu}
\altaffiltext{2}{School of Physics and Astronomy, University of Minnesota, Minneapolis, MN 55455}
\altaffiltext{3}{Department of Physics, UNIST, Ulsan, 689-798, Korea; ryu@sirius.unist.ac.kr}


\begin{abstract}

We analyze data from high resolution
simulations of the generation of compressible, magnetohydrodynamic (MHD) turbulence with properties chosen to resemble conditions in galaxy clusters. In particular, the flow is driven to have turbulence Mach number $\mathcal{M}_t \sim 1/2$ in an
isothermal medium with an initially very weak, uniform seed magnetic field ($\beta = P_g/P_B = 10^6$).
Since cluster turbulence is likely to result from a mix of sheared (solenoidal) and compressive forcing processes, we examine the distinct turbulence properties for both cases. In one set of simulations velocity forcing is entirely
solenoidal ($\nabla\cdot \delta {\vec u} = 0$), while in the other it is entirely compressive ($\nabla\times \delta {\vec u} = 0$).
Both cases develop a mixture of solenoidal and compressive turbulent motions, since each generates the other. The development of compressive turbulent motions leads to shocks, even when the turbulence is solenoidally forced and subsonic.  Shocks, in turn, produce and amplify vorticity, which is especially important in compressively forced turbulence. To clarify those processes we include a pair of appendices that look in detail at vorticity evolution in association with shocks.
From our simulation analyses we find that magnetic fields amplified to near saturation levels in predominantly solenoidal turbulence can actually enhance vorticity on small scales by concentrating and stabilizing shear. The properties, evolution rates and relative contributions of the kinetic and magnetic turbulent elements depend strongly on the character of the forcing. Specifically, shocks are stronger, but vorticity evolution and magnetic field amplification are slower and weaker when the turbulence is compressively forced. We identify a simple relation to estimate characteristic shock strengths in terms of the turbulence Mach number and the character of the forcing.
Our results will be helpful in understanding flow motions in galaxy clusters.

\end{abstract}


\section{Introduction}
\label{intro}

Most of the baryonic matter in galaxy clusters and cosmic filaments exists in the form of very diffuse plasma.   These media, both the hotter, intracluster medium (ICM) and cooler, Warm-Hot Intergalactic Medium (WHIM){\footnote{For simplicity, hereafter, both media will be labeled ``ICMs''.}, are very dynamical environments with active ``weather'' driven by ongoing accretion, substructure motions, occasional, violent merger activity  and at times large energy inputs from starburst-driven galactic winds and very fast outflows from active galaxies (AGNs) \citep[][and references therein]{bj14}. Simulations predict, consequently, that ICMs contain large-scale flows (``ICM winds'') at a fair fraction of the local sound speed, and both simulations and observations indicate that they contain weak-to-moderate-strength shocks, contact discontinuities (known in clusters as ``cold fronts'') and strong bulk shear.  Such flows should become turbulent.  ICM turbulence is manifest in simulations of cosmic structure formation \citep[e.g.,][]{ryu03,fujita04,vazza09,rusz11,xu11,min14, vazza14} and there is observational evidence in clusters that also supports this \cite[e.g.,][]{chur04,shueck,bona10,chur12}. An important feature of the events driving ICM turbulence is that they involve both strong shear and strong compressions that are not necessarily coincident.  Accordingly, ICM turbulence characteristics are unlikely to be uniform on cluster scales.

ICM turbulence has important consequences, including non-thermal pressure 
support, entropy and metal redistribution and probably cosmic ray acceleration. 
In addition, since the ICMs are highly ionized and conducting,
such turbulence may amplify even very weak seed magnetic fields through the small-scale, turbulent 
dynamo \citep[e.g.,][]{schek05,ryu08,cho14}. Those magnetic fields, in turn, will determine important microphysical properties of the ICM, including electrical and thermal conductions and kinetic viscosity that are dependent on the field structure and strength \citep[e.g.,][]{nar01,kunz12,parrish12,gasp13,kunz14,howes15}.  The possible origins and 
structures of ICM seed fields, are varied. There are many candidate sources, ranging
from primordial to plasma-physical, ICM- and galaxy--based \citep[e.g.,][for reviews]{widrow,ryu12}. In 
most cases 
the distributed seed field strength available for amplification in the ICM should be several orders of magnitude less than the $\mu$Gauss field values
inferred to be in clusters from observations \citep[e.g.,][]{bona10}.
Recent Planck observation of cosmic microwave background (CMB) anisotropies, for instance, put a constraint on the upper limit of the primordial field strength, $B \la$ a few nGauss at a scale of $\sim 1$ Mpc \citep{planck15}.
Fermi observation of TeV Blazars, on the other hand, put a lower bound of $\sim 10^{-16}$ Gauss again at a scale of $\sim 1$ Mpc for the void magnetic field strength \citep[e.g.,][]{nv10,chen2014}. The specific origin, strength and distribution of seed fields are beyond the scope of our work here. Once kinetic turbulence develops on small scales, the initial amplification of weak seed fields is exponential and rapid \citep[e.g.,][]{schek05,cho09}, so memory of the initial field properties is largely lost.

While the ICM magnetic fields are very important on small scales, we note that $\mu$Gauss field strengths are dynamically unimportant on large scales in the ICM, since the associated, characteristic Alfv\'en speeds, $v_A =\sqrt{2 P_B/\rho} \la 100$ km/sec, while characteristic ICM sound speeds, $c_s \sim 10^3$ km/sec. As noted, large scale flow motions, and, indeed turbulent velocities are expected to be significant fractions of the sound speed.

Turbulent motions in compressible media will generally include both solenoidal 
($\vec{\omega}=\nabla\times\vec{u}\ne 0$) and compressive 
($(1/\rho)d\rho/dt = -\nabla\cdot\vec{u} \ne 0$) components, where $\vec u$ is velocity, $\vec{\omega}$ is vorticity and $\rho$ is mass density. 
In subsonic turbulence, which should be the most prevalent in the ICM, the compressive motions are often ignored. 
However, since even solenoidal turbulence produces pressure fluctuations $\sim \rho u^2$ \citep{batch51}, there will be density fluctuations if the sound speed is finite, thus, a compressive component to the turbulence \cite[e.g.,][]{kida90,port92,port02,cholaz02,gasp14}. When the turbulence velocities become more than $\sim 10$ \% of the sound speed, these will lead to shocks \cite[e.g.,][]{kida90}.  If the turbulence is forced by compressive motions, then that compressive component typically dominates, even if the turbulent velocities are subsonic \cite[e.g.,][]{fed11,kons12}. In fact, shocks are common in the ICM with Mach numbers up to a few, so some contributions from compressive forcing are likely. It is important to understand these relationships
when discussing ICM turbulence.

Turbulent dynamo amplification of the ICM magnetic field also depends on the character of the turbulence, since amplification comes mostly from stretching via solenoidal motions. Because stretching rates in solenoidal turbulence are generally fastest on the smallest non-dissipative scales, efficient amplification of weak magnetic fields requires that the solenoidal motions driven on large scales cascade to much smaller scales before they are dissipated (call those lengths ``$\ell_d$''). That is, the hydrodynamical Reynolds number, $R_{e,H}$, of the turbulent flow, $R_{eH}\sim u_e L_e/u_d \ell_d \sim (L_e/\ell_d)^{4/3}$, must be large, where $u_e$ and $L_e$ are the fluid velocity and scale of the largest eddies, $u_d$ is the solenoidal velocity on the scale $\ell_d$, and  Kolmogorov velocity scalings, $u_d \sim u_e (\ell_d/L_e)^{1/3}$, were used to obtain the final relation. 

ICM plasmas, being very hot and diffuse, are only weakly collisional, in the sense that Coulomb scattering mean-free-paths, $\lambda_C$, often exceed 10 kpc \citep[e.g.,][]{schek06}, so are large compared to important dynamical scales. If Coulomb collisions were responsible for turbulent kinetic energy dissipation in the ICM, i.e, if $\ell_d \sim \lambda_C$, then from the above relations, and assuming turbulence with $L_e \sim100$ kpc, we would expect $R_{e,H} \la 100$, which is probably too small to allow significant turbulent amplification of ICM magnetic fields. On the other hand, the presence of a weak magnetic field enables a host of plasma-scale instabilities, such as the firehose and mirror instabilities, that very likely reduce the kinetic energy dissipation scales well below $\lambda_C$ \citep[e.g.,][]{schek06,howes08,kunz14,howes15}. Thus, it is probable that ICM $R_{e,H} \gg 100$, as needed to facilitate turbulent magnetic field amplification. Direct ICM kinetic turbulence measurements on scales below a few kpc needed to confirm this explicitly do not yet exist. There are, on the other hand, recent X-ray-based measurements of density fluctuations in several ICMs that are consistent with Kolmogorov, inertial turbulent velocity spectra below 10 kpc \citep{gasp13,zhur15}, seemingly requiring kinetic energy dissipation scales below that, and potentially well below that. Our simulations described below are motivated by the expectation that, indeed $\ell_d$ is commonly sub-kpc, so that turbulence with $R_{e,H}\gg 100$ is also common.

In this same context we mention that, of course, resistive dissipation, or magnetic diffusion, also plays an important role in determining the ability of turbulent flows to amplify magnetic fields. Obviously, if magnetic diffusion is faster than field line stretching on the smallest scales of the kinetic turbulence, magnetic field amplification is not effective. This comparison is generally expressed in MHD through the magnetic Prandtl number of the fluid, $P_{r,m} = \nu/\eta \sim \tau_{d,m}/\tau_{d,h}$, where $\nu$ and $\eta$ are the fluid kinetic viscosity and resistivity, while $\tau_{d,m}$ and $\tau_{d,h}$ represent characteristic magnetic and hydodynamical dissipation time scales. If both viscous and resistive dissipation in ICMs were determined by (infrequent) Coulomb collisions; that is, if so-called Braginskii MHD  \citep{brag65}, applied, then $P_{r,m}\gg 10^{20}$ \citep[e.g.,][]{schek06}. This is  certainly favorable to turbulent magnetic field amplification, although such an extreme Prandtl number is unlikely. We noted above, for a start, that ICM viscosities are likely to be much smaller than in Braginskii MHD. At the same time the ICM resistivity is likely to be substantially larger than the Braginskii MHD model. On the latter we comment that while no direct constraints on $\eta$ values appropriate to ICM turbulence are available, recent analyses of cluster temperature structures strongly suggest that ICM thermal conduction is suppressed by at least several orders of magnitude compared to Braginskii MHD \citep[][]{gasp13,zhur15}. Since both thermal and electrical transports are controlled by the effective electron mean free path, this implies that the electrical conductivity (resistivity)  is substantially reduced (enhanced) in ICMs over values derived from Coulomb scattering alone. Thus, $P_{r,m}$ in ICMs is likely to be considerably smaller than Coulomb-scattering-based estimates, although there is no basis to suggest it is less than unity. The most important condition in our problem is to have $P_{r,m}\ga 1$ \citep[e.g.,][]{schek07,brand14}, which seems very likely. Our simulations reported here are based on numerical, Euler-limit MHD, where, dissipation is negligible in resolved flows, as outlined in the next section. Consequently, the effective $P_{r,m} \sim 1$. 

An additional and very relevant constraint on the degree of magnetic field amplification is the time available.} An initially weak magnetic field embedded in solenoidal turbulence for at least  a couple of tens of large-scale eddy turnover times can be amplified to strengths approaching energy equipartition with the  {\it solenoidal} motions \citep[e.g.][]{cho09,fed11,beres12}. 
However, the available number of such large-eddy turnovers in clusters, $N_e$, is not generally very large. A simple estimate would be $N_e \sim f_d\times u_e/(H L_e)$, 
where $H \approx 70~{\rm km/sec/Mpc}$  
is the Hubble parameter, $u_e$ 
is the largest eddy velocity, $L_e$ 
is the largest eddy size and $f_d < 1$ is an effective time fraction on which ICM turbulence is strongly driven. For characteristic ICM values $u_{e} \sim {\rm{several}\times 100~{\rm km/sec}}$, $L_{e} \sim 100~{\rm kpc}$, $f_d \sim 0.5$, the expectation for $N_e$ is of order 10 \cite[see, also, e.g.,][]{ryu08,cho09b}. So, while it is likely that kinetic turbulence is well-developed and substantial magnetic field amplification takes place, especially on relatively smaller scales, it is, on the other hand, less likely that clusters and cosmic filaments should develop stationary, {\it fully saturated MHD} turbulence states with energy equipartition between kinetic and magnetic energies on large-eddy scales.

To help explore the physics of the above issues we have carried out two series of high resolution simulations of driven  MHD turbulence in compressible, isothermal fluids.
The simulations all begin with a very weak ($\beta = P_g/P_B = 10^6$), and, for simplicity, uniform seed magnetic field in a stationary
medium.  The initial magnetic field value corresponds roughly to $\mathcal{O}(10)$ nGauss levels
in the ICM context (see the next section), although
evolution of the field  past very early, exponential growth times, is not sensitive 
to this choice. Analogous simulations
initiated with non-uniform seed magnetic fields are discussed elsewhere \citep{cho12,emer15}. Turbulence is then driven towards
equilibrium velocities $\sim (1/2) c_s$, where $c_s$ is the isothermal sound speed (turbulence Mach number $\mathcal{M}_t \sim 1/2$). For characteristic ICM sound speeds, $c_s \sim 10^3$ km/sec, the equilibrium turbulent velocities are then several hundred km/sec.

Two limiting cases for turbulence forcing are presented here.  In one case velocity forcing is entirely
solenoidal ($\nabla\cdot \delta {\vec u} = 0$), so the turbulence is predominantly solenoidal in character as it evolves, and the turbulent dynamo is
relatively efficient. In this case vorticity in the turbulence (or, as it turns out, more meaningfully, the enstrophy, $\epsilon = (1/2)\omega^2$,  is initially amplified by vortex stretching dynamics, as in neutral fluid turbulence. But, as the magnetic field approaches saturation, magnetic tension forces reduce vortex stretching, while they concentrate tangential shearing motions inside magnetic structures, and, thereby dominate the (positive) generation of enstrophy in the fully MHD turbulent flows. 
In the other simulation case the forcing is entirely compressive, so there are no applied sources of vorticity. On the other hand, vorticity is seeded and amplified at shocks and by Maxwell stresses, but solenoidal motions remain sub-dominant to the end of the simulations. Then the turbulent dynamo is suppressed. In both cases shocks develop out of the turbulence with strengths that are roughly predictable from standard relations describing density fluctuation amplitudes. While obviously idealized, the two model extremes allow us to identify more clearly the dependencies of the turbulence properties and the resultant magnetic field growth on the nature of the driving. 

Since shocks play such an important role in ICM turbulence, we touch again on the fact that ICMs are weakly collisional plasmas, while our simulations are based on numerical approximations to MHD.  Internal structures of collisionless shocks depend on detailed plasma processes \citep[e.g.,][]{wilson07}. Yet, over sufficiently long time intervals across the full transition they necessarily satisfy the same jump conditions as MHD shocks, since the jump conditions are derived from the basic conservation laws. Similarly, our numerical MHD shocks contain structures, which, in this case depend on the numerical method. In order to evaluate potential influences of internal shock structures in our study we carried out a Favre filtering analysis of our MHD simulations and compared those to analytic assessments of the roles for shocks in similar flows. We found, in fact, good consistency between the methods, strengthening the reliability of our results in the context of ICM shocks.

We present and compare these two studies here. The plan of the paper is as follows. Section 2 outlines our numerical methods and simulation parameters. In \S 3 we discuss necessary physics of vorticity (or enstrophy) and magnetic  field amplification. Section 4 summarizes results of our analysis, while our conclusions are summarized in \S 5. Appendix \ref{shockvort} presents an analytic discussion of vorticity generation across shocks in order to clarify the physical basis of our results, Appendix \ref{filter} looks at the same issue in terms of a Favre-filtered flow analysis of our simulation results, while Appendix \ref{shockfinder} outlines a novel scheme to identify and characterize shocks in grid simulations. That method is used here to compute area-weighted shock Mach number probability distributions.


\section{Numerical Details}
\label{methods}

The simulations were carried out with an isothermal TVD MHD code, updated for performance and parallel scaling from the code presented in \cite{kr01}.
The code uses constrained transport \citep{dai98,ryu98} to maintain a solenoidal magnetic field ($\nabla \cdot B = 0$). 
 It does not explicitly model viscous and resistive dissipations.
The cubic simulation box had dimensions $L_0 = L_x = L_y = L_z = 10$  code units with periodic boundaries. 
Initially the medium was all at rest and  had a uniform density, $\rho_0 = 1$, gas pressure, $P_{g,0} = 1$ (so isothermal sound speed, $c_s = 1$) and a uniform magnetic field in the $\hat x$-direction with $\beta_0 = P_{g,0}/P_{B,0} = 10^6$, so an Alf\'ven velocity, $v_A = \sqrt{2P_B/\rho}\approx 1.4\times 10^{-3}$. 
With these inputs the box sound crossing time is 10 code units. Since our turbulence
velocities become comparable to the sound speed, and the energy containing scale is about 2/3 the box size (see below), we would expect a characteristic
time for turbulence development to be $\mathcal{O}(10)$ in these units, as it was. 

Although we emphasize that our intent has been to make these simulations as scale free as possible, it can be helpful to some readers to have some appropriate characteristic ICM scales in mind. In that spirit,  if we imagine that $L_0 \sim 100$ kpc, while $c_s \sim 10^3$ km/sec, then the equivalent sound crossing time would be $\sim 100$ Myr; i.e., a unit of time in these simulations would then roughly correspond to 10 Myr. Similarly, the ICM gas pressure, $P_g = c_s^2 \rho \sim 2\times 10^{-11}~n_{e-3}$ dyne/cm$^2$, where $n_{e,-3}$ is the electron density compared to $10^{-3}~{\rm cm}^{-3}$. The initial magnetic field with $\beta_0 = 10^6$ would correspond to $B_0 \sim 2\times 10^{-8}$ Gauss, while the strongest ending RMS field values in these simulations (Model S2K) with $\langle \beta \rangle \approx 0.055$ would correspond to $B_{RMS} \sim 5~\mu$Gauss.

As noted in the Introduction, ICM turbulence is expected to be driven by accretion, mergers, and substructure motions, as well as galactic winds and AGN outflows \citep[see, e.g.,][]{bj14}.
\citet{ryu08}, for instance, proposed a scenario in which turbulence is initiated by the vorticity generated at shocks formed during the formation of the large scale structure.
Yet, defining the nature of the source of turbulence is not trivial.
Here we adopt the common approach where turbulence is driven on scales comparable to the box size and the resulting turbulence developed on smaller scales is examined.
Turbulence in our simulations was driven with a method similar to that of \citet{st99} and \citet{ml99}. Velocity forcing $\delta{\vec f}$ was drawn from a Gaussian random field with a power spectrum,  $P_k \propto k^6 \exp(-8 k / k_{exp})$ in the interval $1\le k/k_0 \le 10$,
where $k_{exp}/k_0 = 2$, with $k_0 = 2 \pi / L_0$;
$\delta{\vec u} \equiv \delta{\vec f}\ \Delta t$ was added on intervals, $\Delta t = 0.01 L_0/c_s$.
The forcing power spectrum peaks around $k_d \approx (3/2) k_0$, or
around a scale, $L_d \approx (2/3) L_0$.  With $L_0 \sim 100$ kpc, for instance, $L_d \sim 67$ kpc, which is of order of the scale height of a cluster core.
Forcings have random phases, so they are temporally uncorrelated.
The amplitude of the perturbations was tuned to produce
$u_{RMS} = \langle u^2\rangle^{1/2} \sim 1/2$ (sonic Mach number
$\mathcal{M}_t = u_{RMS} /c_s \sim 1/2$) 
at saturation.
In practice the equilibrium, total turbulent kinetic-energy-equivalent velocities fell in the range  $0.4 \la u_{RMS}\la 0.6$, depending
on the nature of the driving defined below and on the resulting final magnetic field strength. It will be convenient below to refer times to an effective turbulence driving time, $t_d = L_d/u_{RMS}$. In these simulations, $t_d \sim (4/3)L_0/c_s  \sim 13$ (in code units).

 ICM turbulence should derive from dynamics that lead to a mix of solenoidal and compressive driving conditions.  We idealized that here by
using a Helmholtz decomposition of the  driving velocity field. It was thereby separated into solenoidal ($\nabla\cdot\vec{\delta u} = 0$)
and compressive ($\nabla\times\vec{\delta u} = 0$) components.  The fraction of the total driving kinetic energy put into solenoidal motions is designated below by the symbol, $f_s$. 
Results are  presented here for {\it purely solenoidal,} ${\underline{ f_s = 1}}$, and {\it purely compressive,} ${\underline{ f_s = 0}}$, driving. 
The simulations were carried out in each case over a wide range of grid resolutions, $\Delta x = L_0/N$. For the $f_s = 1$ case we show results from simulations carried out on $N^3=1024^3$ and $N^3=2048^3$ cell grids.  For compressive driving, $f_s = 0$, we include results from simulations on $N^3=512^3$ and $N^3=1024^3$ cell grids. 

The so-called integral scale of the velocities developed in the flows, $L_{I,u}$, is a standard tool to help characterize turbulence properties. Specifically,
\begin{equation}
L_{I,u} = \frac{1}{2\pi}\left(\sum_{k=k_0}^{\infty} P_{k,u}dk \right)^{-1}\sum_{k=k_0}^{\infty} P_{k,u}k^{-1} dk,
\label{eq:iscale}
\end{equation}
where $P_{k,u}$ is the power spectrum for the three-dimensional (3D) velocity field. The integral scale, $L_{I,u}$ in the fully-developed, driven turbulence will typically be comparable to, but a bit smaller than the driving scale, $L_d$, mentioned above. For the simulations reported here $L_{I,u} \sim 0.7 L_d$ (see Table \ref{table1}). A characteristic eddy turnover time at the integral scale can then be defined as $t_{e, I} = L_{I,u}/u_{RMS} = (L_{I,u}/L_d) t_d$.  In our simulations $t_{e,I} \sim 10$ (in code units).

As mentioned above, all of the numerical models presented here were computed using a version of the isothermal ($P_g \propto \rho$) MHD TVD code in which explicit viscosity and resistivity are absent; the simulations were not intended to resolve the viscous or resistive dissipation scales of the flows. They employ, instead, nonlinear switches that are sensitive to variations in fields at the limit of what can be resolved on the computational grid and apply minimal damping needed to stabilize numerical instabilities at Euler discontinuities including shocks, contact discontinuities and slip surfaces. For this reason such simulations are sometimes called ``Implicit Large Eddy Simulations'' (ILES){\footnote{Also see, e.g.,  \citep[][]{aspden08} for further discussion of ILES.}; they implicity assume sub-grid structures corresponding to discontinuities (see \S 1 and Appendix B for further comments on this issue)

Although continuously driven simulations of this kind develop a Kolmogorov-like inertial range, they do not have physically defined kinetic or resistive dissipation scales, such as those defined in Navier-Stokes flow. Thus, it is not possible to establish corresponding physical hydrodynamical and magnetic Reynolds numbers or, for that matter the magnetic Prandtl number. However, the simulations do have well-defined dissipation scales related to the mesh scale. Peak to trough variations are typically spread over about $4\Delta x$, corresponding to a dissipation length, $\ell_d \sim 4\Delta x \equiv 4L_0/N$. This grid-based dissipation scale is approximately the same for all fields. Hence, the effective Prandtl number is effectively $P_{r,m} \sim 1$. All the presented models exhibit inertial-like power spectra when turbulence first develops (see Figures \ref{sol:spec} and \ref{comp:spec}), but before magnetic stresses begin to influence the flows. Thus, we can construct estimates for effective hydrodynamical Reynolds numbers of the turbulence based on the integral scale defined in equation (\ref{eq:iscale}) as $R_{e,H} = (L_{I,u}/\ell_d)^{4/3}$, where the 4/3 index assumes for simplicity Kolmogorov velocity scaling. Reynolds numbers computed in this fashion range between about 240 and 1390 for the simulations presented here.

Table \ref{table1} summarizes the basic inputs and characteristics of four models discussed in the following sections.


\section{Evolution of Vorticity and Magnetic Fields}

\subsection{Vorticity and Enstrophy Generation}
\label{vorticity}

Vorticity is a key measure of solenoidal turbulence, since solenoidal turbulence always includes circulation, and vorticity provides a measure of eddy circulation rates.  It is important, however, to keep in mind that solenoidal turbulent energy is generally concentrated on largest-eddy scales, ($\delta u_{\it{l}}^2 \propto {\it{l}}^{2/3}$ for Kolmogorov turbulence) while the associated vorticity generally cascades to smaller scales closer to dissipation scales ($|\omega_{\it{l}}|\propto{\it{l}}^{-1/3}$ in the above case). This difference will be important in our discussion of results in \S 4. To streamline our discussion there we briefly review some of the basics of its generation and amplification. The following equation governing vorticity is obtained from the curl of the Navier Stokes
equation with magnetic (Maxwell) stresses, $\vec j\times\vec B$, added; 
\begin{equation}
\frac{\partial \vec \omega}{\partial t} =- \nabla\cdot(\vec u\vec\omega) + (\vec\omega\cdot\nabla)\vec u +\frac{1}{\rho^2}\nabla\rho\times\nabla P_T+ \nabla\times\left(\frac{\vec T}{\rho}\right) + \nu\left(\nabla^2\vec\omega +\nabla\times \vec G\right) + \nabla\times\vec a_{drv} ,
\label{eq:vort}
\end{equation}
where $\vec a_{drv}$ is the impulsive driving force, $P_T = P+P_B$ is the sum of the gas pressure, $P$, and the magnetic pressure, $P_B = (1/2)B^2$ in the units we use here.  $\vec T = \vec B\cdot\nabla\vec B$ is
the magnetic tension,
$\nu$ is the kinematic viscosity (assumed constant), while $\vec G = (1/\rho)\nabla\rho\cdot\vec{\vec S}$, with $\vec{\vec S}$ the standard traceless strain tensor \cite[e.g.,][]{mee06}. 
The box-averaged mean, $\langle\partial\vec\omega/\partial t\rangle = 0$, since forcing is uncorrelated across the box in these simulations ($\langle\nabla\times\vec a_{drv}\rangle=0$).
We note for clarity that the first two RHS terms can be combined to produce the more commonly used term, $\nabla\times(\vec u\times\vec \omega)$.

The first RHS term in equation (\ref{eq:vort}) accounts for conservative advection of the vector vorticity, while the second represents vortex stretching.  The remaining terms are vorticity source or sink terms.
Our calculations simulate isothermal
fluids, so $P\propto \rho$ and the baroclinic source term in equation (\ref{eq:vort}), $(\nabla\rho\times \nabla P)/\rho^2$, vanishes everywhere.  On the other hand, the magnetic pressure will not generally be a barotropic function of density; that is, $P_B \ne P_B(\rho)$, so the magnetic field can contribute a quasi-baroclinic  source term even in isothermal flows. However, magnetic pressure variations and density variations turn out to be strongly anti-correlated in turbulent flows\footnote{This behavior reflects the well-known dominance of slow mode oscillations over fast modes in compressible MHD turbulence \cite[e.g.,][]{kowlaz10}.}, so their gradients are, after all, nearly anti-parallel. This term is then mostly sub-dominant in our simulations, even when the magnetic field is not weak. The magnetic tension term in equation (\ref{eq:vort}) represents the resistance of the field to bending and stretching, which inhibits solenoidal motions, including those that promote vortex stretching. The magnetic terms in equation (\ref{eq:vort}) are initially small in our simulations, because the initial magnetic field is weak and uniform. But, as we shall see, they have roles to play once motions distort the field, and especially if the magnetic field is strongly amplified by dynamo action. The final, non-forcing term in equation (\ref{eq:vort}) accounts for viscous dissipation.
Our simulated fluids are nominally ``ideal'', since, formally, $\nu = 0$. On the other hand, numerical dissipation mimics viscosity on scales
close to the grid resolution, producing effects resembling the viscous term in equation (\ref{eq:vort}).  It allows viscous dissipation, shock formation and, through the viscous stress ($\nu\nabla\times \vec G$) term, vorticity creation in shocks. Appendix \ref{filter} examines vorticity evolution in mass-weighted Favre-filtered ideal flow, where we obtain an analogous Reynolds-stress source term in ideal flows. 

We emphasize that shocks are expected, do develop in our simulated turbulent flows and are important, even though the turbulence is subsonic. This will be discussed in detail in \S 4.1.2 and \S 4.2.2. Turbulent motions produce those shocks. The shocks, in turn, generate vorticity. Vorticity changes across shocks are most simply understood by looking explicitly at shock jumps. Several authors have explored this issue analytically \cite[e.g.,][]{crocco,kev09}.  Since shocks turn out to be a principal generator of vorticity in our initially irrotational, compressively driven turbulence simulations, and since the full physics of shock generation of vorticity is not necessarily intuitive, we present an outline of the relevant processes in Appendix A, extended to include magnetic field influences. The basic consequent hydrodynamical behaviors are contained in equation (\ref{delomega}); magnetic field augmentations are contained in equation (\ref{delomegab}). As an additional approach to understanding shock generation of vorticity, our Appendix \ref{filter} examines these interactions in terms of filtered flows, which largely eliminates numerical artifact issues at shock discontinuities in the simulations. From that analysis we find, in fact, that the unfiltered, raw analysis of our simulations gives reliable results. From this we conclude in the context of ICM shocks, which involve weakly collisional plasmas, that the key physics associated with the vorticity evolution at shocks derives from conservation of mass and momentum, so is independent of the specific dissipation physics.

While equation (\ref{eq:vort}) provides an effective description of local vector vorticity evolution, $\vec\omega(t,\vec x)$, the vector field is not very useful for an understanding of the global vorticity evolution in our periodic-box simulations, since $\langle\vec\omega(t)\rangle = 0$. The vorticity magnitude, $|\omega|$, or more conveniently, the enstrophy, $\epsilon=(1/2)\omega^2$, does provide a useful measure of global evolution. The dot product of equation (\ref{eq:vort}) with the vorticity, $\vec\omega$, gives an enstrophy equation,
\begin{equation}
\frac{\partial\epsilon}{\partial t}=F_{adv} + F_{stretch} + F_{comp} + F_{baroc} + F_{mag} +  F_{diss},
\label{eq:enst}
\end{equation}
where we have combined and arranged RHS ``flux'' and source terms to emphasize their physical interpretation\footnote{We drop forcing contributions in equation (\ref{eq:enst}), since, the velocity forcing, $\vec a_{drv}$, (which is only on large scales) is essentially uncorrelated with $\vec\omega$. The net forcing contribution to $\partial\epsilon/\partial t$, while not exactly zero, is negligibly small.}. Specifically,
\begin{eqnarray}\label{eq:enstsource}
F_{adv} =- \nabla\cdot (\vec u \epsilon) =-( \epsilon\nabla\cdot\vec u + \vec u\cdot\nabla\epsilon),\nonumber\\
F_{stretch} = \vec\omega\cdot (\vec\omega\cdot\nabla)\vec u =  2\epsilon (\hat\omega\cdot\nabla)\vec u \cdot \hat\omega,\nonumber\\
F_{comp} = -\epsilon \nabla\cdot \vec u = \frac{\epsilon}{\rho}\frac{d\rho}{dt} = -\nabla\cdot (\vec u \epsilon) + \vec u\cdot\nabla\epsilon,\\
F_{baroc} = \frac{\vec\omega}{\rho^2} \cdot (\nabla\rho\times\nabla P),\nonumber\\
F_{mag} = \frac{\vec\omega}{\rho^2} \cdot \left(\nabla\rho\times\nabla P_B\right) + {\vec\omega} \cdot \nabla\times\left(\frac{\vec T}{\rho}\right),\nonumber\\
F_{diss} = \nu\vec\omega\cdot\left(\nabla^2\vec\omega + \nabla\times \vec G\right),\nonumber
\end{eqnarray}
where $\hat\omega$ is the unit vector in the direction of $\vec \omega$.

The enstrophy advection term, $F_{adv}$ is conservative, so that its integral over our periodic simulation volume always vanishes. Similarly, in our simulations the baroclinic term, $F_{baroc} = 0$. The magnetic term, $F_{mag}$, is, as noted above, initially very small, because of the initially weak and uniform magnetic fields. In the solenoidal driving simulations, however, this term eventually plays a significant role in the evolution of the enstrophy and the turbulent energy. $F_{mag}$, being dominated by magnetic tension, generally reduces the rate of vortex stretching. It can, however, also concentrate and stabilize tangential shear on small scales, so lead to a net increase in enstrophy above the neutral fluid level. Recall from above that in Kolmogorov turbulence $\epsilon_{\it{l}} \propto {\it{l}}^{-2/3}$, so that the enstrophy is concentrated on small scales even without this influence from the magnetic tension. The remaining contributors from equation (\ref{eq:enst}) to enstrophy increases in our simulations are the $F_{stretch}$ and $F_{comp}$ terms that account for vortex stretching and net enstrophy production in compressions, respectively. Note that the fluid compression rate, $-\nabla\cdot\vec u$, enters into both of $F_{adv}$ and $F_{comp}$ terms. However, whereas $F_{adv}$ always integrates to zero in a periodic domain, $F_{comp}$ does not, if there is a net alignment of the velocity and enstrophy gradient fields, so that  $\int F_{comp}~dV = \int \vec u\cdot\nabla\epsilon~dV > 0$. This alignment is usually true in shocks (see Appendices \ref{shockvort} and \ref{filter}), so this term turns out in our isothermal flows to  represent the dominant contribution of shocks to positive enstrophy growth. As we see below, the $F_{comp}$  term is especially important in the compressively forced turbulence, but less so in the solenoidally forced turbulence. The dissipative term, $F_{diss}$, should also be present through numerical viscosity, although we find in our simulations that the other terms alone can account reasonably well for the enstrophy evolution in these simulations.

\subsection{Magnetic Field Amplification}
\label{magfield}

The magnetic field evolution is governed by the induction equation, whose structure, of course, is very similar to equation (\ref{eq:vort}) governing vorticity.  For a generalized Ohm's law in the MHD approximation the induction equation  is \citep[e.g.,][]{kuls97,boyd},
\begin{equation}
\frac{\partial \vec{B}}{\partial t} = \nabla\times(\vec{u}\times\vec{B}) + \eta\nabla^2\vec{B}-\frac{c}{\sqrt{4\pi}e n_e^2}\nabla n_e\times \nabla P_e, 
\label{eq:induct}
\end{equation}
where $\eta$ is the resistivity  (assumed constant and isotropic). The last term
on the right in equation (\ref{eq:induct}), the so-called ``Biermann battery'' source term, 
is analogous to the baroclinic vorticity source term, and
comes from different electron and ion mobilities\footnote{Note that $\sqrt{4\pi}$ appears here, since we chose to express the magnetic field in units such that $P_B = (1/2)B^2$.}. Here, $n_e$ and $P_e$ are the electron density and pressure, respectively. It could be important as a seed magnetic field generator in the early universe; the resulting magnetic field would be weak, for instance, $\left<B\right> \la 10^{-20}$ G in proto-clusters \cite[e.g.,][]{kuls97,ryu12}.  We do not consider creation of magnetic fields in this work,  only evolution of existing magnetic fields, so neglect that term.

Similar to the case for vorticity, $\vec\omega$, equation (\ref{eq:induct}) for the vector magnetic field, $\vec B$, is not very useful for study of the global evolution of the magnetic field in our simulations. Analogous to that case, however, we can construct an equation for magnetic pressure (or equivalently energy density), 
\begin{equation}
\frac{\partial P_B}{\partial t} = -\nabla\cdot(\vec u P_B) + 2 P_B(\hat B\cdot\nabla)\vec u\cdot\hat B - P_B\nabla\cdot\vec u + \eta\vec B\cdot\nabla^2\vec B,
\label{eq:magpress}
\end{equation}
where we have dropped the Biermann source term, and $\hat B$ is the unit vector in the direction of $\vec B$. Analogous to equation (\ref{eq:enst}), the first RHS term in equation (\ref{eq:magpress}) represents advection of magnetic energy and integrates to zero over our periodic simulation boxes. The second RHS term is the magnetic field analogy to the vortex stretching term. It is the essential contributor to magnetic energy growth in the turbulent dynamo. Taken by itself in steady solenoidal turbulence without any back-reaction, it leads to an exponential growth in the magnetic pressure, with a rate, $\Gamma \sim |\hat B\cdot\nabla\vec u| \sim u_{\it{l}}/{\it{l}}\sim \omega_{\it{l}}$, with $\it{l}$ an effective scale over which magnetic flux tubes are stretched by velocity fluctuations. The third term is analogous to the $ F_{comp}$ term in equation (\ref{eq:enst}), while the final term
represents resistive dissipation of magnetic energy.
Again our simulations are assume to be ``ideal''; \ie formally that $\eta = 0$. However, once
again, numerical approximations lead to diffusion and dissipation of
magnetic fields on grid resolution scales, so mimic the effects of a finite $\eta$. Since both the numerical resistivity and the numerical
viscosity represent dissipation on the same, grid resolution scale, the effective magnetic Prandtl number, as discussed in \S 2, should be $P_{r,m} = \nu/\eta \sim 1$.

Our initial magnetic field, with $P_B = 10^{-6} P_g$, is too weak to have any significant
dynamical influence at the start of these simulations. However, through the kinematic fluctuation dynamo
the field is quickly amplified once solenoiodal turbulent motions have developed on
small scales.  Once the dynamo is underway the field energy should
be amplified exponentially ($E_B \propto \exp(\Gamma t)$) on a timescale $ \Gamma^{-1} \sim  t_m = {\it{l}_m}/u_m \sim 1/\omega_m$, where $\it{l}_m$ is the minimum length scale of solenoidal velocity fluctuations and
$u_m$ is the associated velocity fluctuation. Following \cite{beres12} to allow for turbulent flux diffusion, we can estimate more quantitatively the expected  exponential-phase growth rate of the magnetic energy to be $\Gamma \approx 0.6/t_m$.  Once the magnetic tension on lengths $\it{l}_m$, $T_{\it{l_m}} \sim P_{B_{\it{ l}_m}}/\it{l_m}$,  is
comparable to Reynolds stresses on those scales; i.e., when $P_{B_{\it{l}_m}}/{\it{l}_m} \sim \rho u(l)^2/{\it{l}_m}$ or when $E_B({\it{l}_m})\sim E_K({\it{l}_m})$,  magnetic field growth on those scales will saturate as
it back-reacts on the motions and reduces vortex and magnetic field stretching. We will see this directly in \S 4.1.1. The field can continue to be amplified by vortex stretching motions on larger scales until it saturates there for similar reasons. For Kolmogorov scaling the kinetic energy described by this evolution
scales as $E_K({\it{l}}) \propto {\it{l}}^{2/3}$. Since, with this scaling, the eddy turnover time is $t_e =  {\it{l}}/u(l) \propto {\it{l}}^{2/3}$, the magnetic energy
during this phase is expected to show a linear growth in time, $P_B \propto t$.  During this linear growth period the energy containing scale of the magnetic field should increase with time as ${\it{l}_B} \propto t^{3/2}$ (again assuming Kolmogorov scaling).  This phase continues until the scale reaches the turbulence driving scale, $L_d$, actually $\sim (1/2 - 1/3) L_d$ \citep[e.g.,][]{cho09b}. By the time the growth in magnetic energy on large scales is balanced by Ohmic dissipation on small scales, roughly as  $E_B$ approaches the kinetic energy contained in solenoidal turbulent motions. How long that takes depends on the nature of the turbulence, but, saturation from an initially weak field will generally span many large-scale eddy turnover times.

The expected magnetic field evolution in the compressively forced situation is different from the solenoidally forced situation in important respects. First, since small scale dynamo amplification depends on solenoidal motions, and those are much reduced in that case, the rate of magnetic energy growth will be much slower. Field amplification will also saturate at much lower energy levels, since the available energy to amplify the field is reduced, as well.


\section{Discussion of Results}


\subsection{Purely Solenoidal Forcing: $f_s = 1$}
\label{results:sol}

We look first at the evolution of the turbulence developed from purely solenoidal forcing, $f_s = 1$, starting with an overview of the various turbulent energy components and the enstrophy, then examine the properties of associated density fluctuations and shocks found in our simulations.

\subsubsection{Energy and Enstrophy Evolution}

Figure \ref{sol:evol} illustrates the kinetic
and magnetic energy evolution in simulations with two different resolutions (models S1K and S2K). 
The early kinetic turbulence evolution is essentially hydrodynamic, since, as discussed in \S3, the initial
magnetic field is too weak to modify motions significantly. Our setup gives a characteristic timescale on the driving scale,
$L_d \approx 6.7$, of $t_d = L_d/u_{RMS} \sim 13$. Solenoidal kinetic energy cascades
to smaller scales on an eddy turnover time, $t_{\it{l}} \approx \it{l}/u_{\it{l}}\sim 1/\omega_{\it{l}}$, so hydrodynamical, solenoidal
turbulence is well developed by $t \sim t_d\sim 1/\omega_{L_d}$ in the $f_s = 1$ simulations. The kinetic
turbulent energy, $E_K$, briefly peaks after one large-eddy turnover near $t \sim 15$ and then levels out by
$t \sim 25$ at $E_K \approx 0.18$, corresponding
to $u_{RMS} = 0.6$.  That behavior obtained for the $f_s = 1$ simulations we did with grid resolutions as coarse as $256^3$. The total kinetic plus magnetic energy remained close to $E_T \approx 0.18$ from $t \sim 25$ forward in all the $f_s = 1$ simulations. Thus, since the total thermal energy is fixed by the isothermal equation of state and does not change in our periodic volume, 
the total energy stored in the system was roughly constant
from $t \sim 25$ onwards. 
The early enstrophy evolution mirrors this behavior, as we would expect and as is illustrated in Figure \ref{soldenst}. The rate of change in the total enstrophy\footnote{The rate of total enstrophy change is obtained from the explicit difference between total enstrophy values in nearby, saved time steps in the simulations.}, $\partial\epsilon/\partial t$, peaks just after $t = 10$, while the cascade of solenoidal kinetic energy to small scales is developing, then adjusts, so that after $t\sim 30$ it remains close to zero. We will look more closely at the enstrophy evolution below.

Figure \ref{sol:evol} shows that the magnetic energy evolution histories are very similar in the two $f_s = 1$ simulations, with a  slightly greater
magnetic field enhancement in the higher resolution simulation reflecting the effectively smaller viscosity in that case. Much of this difference is actually created early on, but after $t\sim 15$, when enstrophy on small scales is fully developed and during the period of exponential magnetic field growth. The difference reflects the fact that the minimum eddy scale, $ {\it{l}_m}$, is larger in the lower resolution, S1K, simulation, so the initial growth rate, $\Gamma \propto {\it{l}_m}^{-2/3}$, is reduced accordingly.  The exponential magnetic energy growth is followed, as expected, by a period of linear energy growth in the interval, $20 \la t \la 80$.
The transition into linear growth begins when $E_B/E_K \sim$ few \% and ends several large-scale eddy turnover times ($\sim t_d \sim 13$) later when $E_B/E_K \sim 30$\%.  Figure \ref{soldenst} shows during this period that the amplification of enstrophy by way of vortex stretching ($F_{stretch}$) is greatly reduced. This is a direct consequence of the inhibition by magnetic tension of stretching motions. Somewhat surprisingly, however, in the same time period there is a comparable {\it increase} in enstrophy amplification due to magnetic tension ($F_{mag}$), so that the total rate of enstrophy amplification remains almost unchanged and balanced by dissipation. We explain that below. By the end of the
S2K simulation  at $t = 130$ the magnetic energy to kinetic energy ratio is $E_B/E_K \sim 1/2$. The magnetic energy is still growing slowly, so over very long times it should eventually approach close to $E_B/E_K \sim 2/3$, the value observed in simulations of incompressible turbulence \citep[e.g.,][]{ryu08,cho09}.

After $t \sim 40$, $E_K$ drops in response to magnetic field back-reaction 
towards an eventual value, $E_K \approx 0.12~(0.13)$, in the S2K (S1K) simulation,
corresponding to $u_{RMS} = \mathcal{M}_t \approx 0.5$.
The solenoidal component contains about 93\% of the kinetic energy from $t \sim 20$ on,
so about 7\% is compressional (not shown in the figure). 
While sub-dominant, compressional motions, including shocks, are present, however, and we discuss their properties below.

Figure \ref{sol:spec} and Figure \ref{solimages} illustrate important changes in distribution properties of kinetic and magnetic energies (and associated stresses) as the magnetic field back-reaction  plays an increasingly significant effect. Figure \ref{sol:spec} shows the kinetic and magnetic energy power spectra of the S2K simulation at $t = 20$ and $t = 130$. At the earlier time the turbulence is still essentially hydrodynamic with an approximately Kolmogorov form,
$E_K(k) \propto k^{-5/3}$, roughly in the interval $2\la k \la 50$, where $k$ is in units of $k_{min} = k_0 = 2\pi/L_0$. On all but very small scales the magnetic energy is much less than the kinetic energy still. By $t = 130$, when $E_B$ is near saturation, magnetic energy is at least comparable to kinetic energy on all but the largest scales. Magnetic tension back-reaction has become important, and substantial kinetic energy has been removed from a wide range of scales below $L_d$, flattening $E_K(k)$ in the process. Note, as part of this, that $E_K(k)$ is not reduced for $k \ga 200$. In fact, at $t = 130$ the enstrophy power ($k^2 E_K(k)$) on these scales actually is increased from what it is at $t = 20$  and is strongly peaked around $k \approx 200$ (not plotted explicitly here). This is a reflection of the previously noted dominant enstrophy amplification after $t\sim 40$  by magnetic tension contributions as seen in Figure \ref{soldenst}.  $F_{mag}$ also is strongest on these scales, as it turns out. Associated changes in the character of the magnetic field are illustrated in Figure \ref{solimages}, which volume renders the magnetic field intensity at the same two times. At the earlier time the strongest magnetic field is broadly distributed into relatively short ``worm-like'' filaments, generated by the cylindrical sheath of high rate of strain common in vortex tubes. By the later time, however, the strongest field structures are longer, but more important to our discussion here, the transverse character of the structures has transformed from round tubes into flattened, filamentary ribbons, as evident in the image. As we will discuss in detail elsewhere \citep{porter15}, the ribbon cross sections consist of laminations separated roughly on the dissipative scale with magnetic field lines along the long axis confining transverse, tangential shear layers. The magnetic tension in the ribbons both confines the shear layers (so enhances enstrophy on those scales) and stabilizes them against Kelvin-Helmholtz instabilities. 

\subsubsection{Compressions and Shock Generation}

As outlined in the introduction, even though these turbulent flows are subsonic and predominantly solenoidal in character,  pressure fluctuations must produce density fluctuations. The statistical properties of the compressions in our $f_s=1$ models are characterized in Figure \ref{sol:pdf}, which shows evolution of
the density probability distribution (PDF) for the S2K simulation at $t = 20, 80, 130$.
The form of the
PDF is approximately the log-normal distribution, originally suggested for supersonic turbulence \citep{vaz94}. The standard deviation of the PDF decreases in this case from
$\sigma \approx 0.19$ at $t = 20$, when the kinetic turbulence Mach number, $\mathcal{M}_t \approx 0.6$, to $\sigma \approx 0.15$ at $ t = 130$, as magnetic field tension
begins to become significant and reduces the kinetic turbulence Mach number to $\mathcal{M}_t \approx 0.5$.
Both results are consistent with the simple scaling relation,
$\sigma^2 = \ln{(1 + b^2 \mathcal{M}_t^2)}$, using $b = 0.3$, in
agreement with $b \approx 1/3$ predictions for solenoidal forcing in compressible, isothermal turbulence \citep[e.g.,][]{fed09,kons12,molina12}.
Thus, characteristic density and gas pressure fluctuations in the
S2K simulation after solenoidal turbulence develops are $\sim 15 - 20$\%.

Shocks form in this case especially by way of collisions between compression waves. The strength of the density fluctuations can then provide a way to estimate a characteristic, expected shock strength. In isothermal flows, shock jump condition is simply,
\begin{equation}
\frac{\delta\rho}{\rho} =\mu = \mathcal{M}_s^2 - 1,
\label{machjump}
\end{equation}
where $\mathcal{M}_s$ is the shock Mach number.
If,  as a crude model, we set $\mu = \sigma$, so match the density PDF standard deviation to a shock jump to define the characteristic shock Mach number, $\mathcal{M}_{s,c}$,
\begin{equation}
\mathcal{M}_{s,c} \approx \sqrt{1+\sqrt{\ln{(1+b^2\mathcal{M}_t^2)}}}~~\frac{\approx}{\sigma<<1} ~~1+\frac{1}{2}\sigma \sim 1+ \frac{b}{2}\mathcal{M}_t.
\label{eq:solmach}
\end{equation}
For our solenoidal simulations with $\mathcal{M}_t = 0.5$ this leads to $\mathcal{M}_{s,c} - 1  \sim 1/12 \approx .08$. Figure \ref{solmachdist} shows the  probability distribution function, $f(\mathcal{M}_s)$, measured at several times for shocks in the S1K simulation obtained using methods introduced in Appendix C. The dotted line in the figure corresponds to an exponential PDF, $f(\mathcal{M}_s) \propto \exp{(-(\mathcal{M}_s-1)/0.08)}$. That is, the measured shock distribution is determined by a characteristic Mach number, $\mathcal{M}_s = 1.08$, very consistent with equation (\ref{eq:solmach}). The total area of the detected shocks with Mach numbers exceeding $\mathcal{M}_{s,c}$ averaged over time is approximately 100 code units ($\sim L_d^2$), corresponding to a mean shock spacing, $\it{l}_s \sim L_d$. Then the simulation box, or more physically, the largest driven eddy, contains, on average, one such shock spanning that eddy scale.


\subsection{Compressive Forcing: $f_s = 0$}
\label{results:com}

\subsubsection{Energy and Enstrophy Evolution}

The evolution and properties of the compressively forced turbulence with $f_s = 0$ have several important differences in comparison to the solenoidally forced turbulence with $f_s = 1$. Figure \ref{comp:evol} illustrates the energy evolution of the two kinetic turbulence components (solenoidal and compressive) along with the magnetic energy for the C1K simulation. In this case the total turbulent energy is almost fully accounted for over the entire simulation time by the compressive kinetic energy, as one might expect. An approximate total and compressive energy steady state condition, with $ E_K \approx 0.08$ ($\mathcal{M}_t = 0.4$) is achieved before $t = 10$, so within one sound crossing time of the simulation box.  However, after $t \approx 5$ there is a rapid growth in solenoidal kinetic energy, and that component saturates at a level  $E_{K,s}/E_{K,c} \sim 1/15$ after $t \sim 20-25$.  Magnetic energy grows in response to the solenoidal motions, but $E_B/E_{K,t}<0.01$ even at the end of the simulation ($t = 120$).

The origins of the solenoidal kinetic energy in this simulation are made clear by examination of Figure \ref{denst} and by noting that significant shocks from collisions of compressively driven motions first form around $t \ga 5$  as in Figure \ref{compfig}.  In the interval $5 \la t \la 10$  the compressive kinetic energy approaches its equilibrium level, and shocks resulting from collisions between compressive waves become common. Figure \ref{denst} shows the volume-integrated enstrophy growth rate, $\partial\epsilon/\partial t$, before $t \approx 16$, along with the volume-integrated enstrophy ``flux terms'', $F_{stretch}$, $F_{comp}$ and $-F_{mag}$\footnote{$F_{mag}<0$, so we reverse its sign in the plot.} in  equation (\ref{eq:enst}) (see also equation (\ref{eq:enstsource})). Before $t \approx 5$ there is a tiny enstrophy seeding, due primarily to numerical truncation. But $\partial\epsilon/\partial t$ increases abruptly around $t = 5$, when shocks first develop, growing almost three orders of magnitude before $t = 10$. The total growth rate is almost perfectly matched during that interval by $F_{comp}$. By $t \approx 10$ the solenoidal kinetic energy, $E_{K,s}$, is approaching its equilibrium level (Figure \ref{comp:evol}) and magnetic tension ($F_{mag}$) and viscous dissipation (numerical, of course) have begun to inhibit enstrophy growth. Consequently, the enstrophy growth rate peaks and  is then somewhat less than $F_{comp}$. In any case, since the integrated $F_{comp}$ comes primarily from shocks (see \S 3.1, Appendix \ref{filter}), it is then clear that enstrophy growth during the early evolution of the compressively forced turbulence is a by-product of shock formation. That point is shown explicitly in Figure \ref{compfig}, which shows zoomed-in images of shock structures and associated enstrophy source terms in a two-dimensional (2D) slice of one of the $f_s = 0$ simulations at $t = 8$.

Since the early enstrophy growth in the $f_s = 0$ simulations is localized to shocks, we would expect magnetic field amplification to be similarly distributed. Indeed, that is the case. Volume renderings of the magnetic field distributions in the C1K simulation are shown in Figure \ref{compimages}. At $t = 20$ the strongest magnetic field is obviously associated with regions of recent shock passage. Even much later, at $t =120$, when both the compressive and solenoidal turbulence motions are near steady-state and broadly distributed, the distribution of magnetic field is obviously clumpy and absent of the ribbon-like topologies noted for the solenoidal case in Figure \ref{solimages}.

The influence of shocks also shows in the turbulent energy power spectra, as shown  for the C1K simulation at two times, $t = 15$ and $t = 120$, in Figure \ref{comp:spec}. The compressive kinetic energy spectra are almost the same at the two times and resembles a Burgers form, $E_{K,c}(k) \propto k^{-2}$, so distinctly steeper than Kolmogorov, roughly over the range $5 \la k \la 100$. The solenoidal kinetic energy spectrum shows modest growth during this interval over all ranges of $k$. It is also distinctly steeper than Kolmogorov, again reflecting the shock origins of these motions.

Since the magnetic field amplification by way of the turbulent dynamo comes  primarily from flux-tube stretching (equation (\ref{eq:magpress})), it is essentially dependent on the solenoidal turbulence component. Because that component in the $f_s = 0$ simulations always represents less than 10\% of the turbulent kinetic energy, it is not surprising that the magnetic field grows much more slowly and remains much weaker than in the $f_s = 1$ simulations. Recall in the S1K simulation that the linear growth of the magnetic field approaches saturation to $E_B/E_{K,s} \sim 1/3$  around $t \sim 80-100$. In contrast, for the C1K, compressive forced simulation, the magnetic field is still in the linear growth phase with $E_B/E_{K,s} \sim 1/8$ ($E_B/E_{K,t} \sim .008$) at the end of the simulation, $t = 150$.  We extended a lower resolution $f_s = 0$ simulation, ChK to $t = 625$, and found even on that much longer time interval that the magnetic energy was still very slowly increasing in time, so had not reached a fully saturated level. Thus, the efficiency of magnetic field energy generation from (total) turbulent energy input depends on the nature of the turbulence forcing.

\subsubsection{Compressions and Shock Generation}

Given the nature of forcing in the $f_s = 0$ simulations and the resulting dominance of compressive kinetic energy, we should expect the density fluctuations and their associated shocks to be stronger than in solenoidally forced turbulence of the same Mach number, $\mathcal{M}_t$. Indeed, as Figure \ref{compdf} shows, the density PDF in our $f_s = 0$ simulations exhibit a log-normal form, just as for our $f_s = 1$ simulations, but with standard deviations, $\sigma$, about twice as large. Again, applying the scaling relation,
$\sigma^2 = \ln{(1 + b^2 \mathcal{M}_t^2)}$,  with $\mathcal{M}_t = 0.4$, we find $b = 0.9$. This is in good agreement with previous prediction of $b \approx 1$ for compressive forcing \citep[e.g.,][]{kons12}. 

As before we can compare this measure of the distribution of density fluctuations to the distribution of shock strengths, which is illustrated in Figure \ref{compmachdist}. Once again the shock PDF is well described by an exponential form; in this case  $f(\mathcal{M}_s) \propto \exp{(-(\mathcal{M}_s-1)/0.125)}$, corresponding to $\mathcal{M}_{sc} = 1.125$. The predicted value from equation (\ref{eq:solmach}) with $b = 0.9$ and $\mathcal{M}_t = 0.4$ is $\mathcal{M}_{s,c} = 1.16$, which, although slightly larger than measured, represents reasonable agreement, given the simplicity of the model. In this case the total area of detected shocks with $\mathcal{M}_s \ge \mathcal{M}_{s,c}$ is about $400 = 4 {L_d}^2$, so about four times greater than for the $f_s = 1$ case. Then the mean separation between such shocks is $\it{l}_s \sim (1/2) L_d$. The higher frequency of moderate strength shocks is consistent with the broader density distribution and the steeper energy power spectra mentioned above.


\section{Conclusion}
Compressible turbulence in ICMs is likely to be common at levels that require consideration of both solenoidal and compressive motions. Both components will be present, since processes that drive ICM turbulence are likely to include both solenoidal and compressive forcing, and  since each kind of turbulence can create the other. Shocks formed in the turbulence are the central link in that process. However, the energy partitioning of these components, as well as their energy spectra, depend on the character of the forces that drive them.
Since ICMs are electrically conducting, solenoidal turbulent motions will, by the turbulent dynamo, amplify magnetic fields that thread them. The rate and effectiveness of that amplification depends on the strength of the solenoidal turbulent component, so, also on how the turbulence is driven. 
 We found that while shocks are stronger, the solenoidal component contains less energy and so magnetic field amplification is less efficient, when turbulence is driven with compressive forcing.

As the magnetic field becomes dynamically important, magnetic tension will inhibit solenoidal motions, leading to anisotropies in the turbulence \citep[e.g.,][]{gold95} and  reductions in kinetic energy. Somewhat ironically, however, the magnetic tension forces can add vorticity on small scales by concentrating and protecting shear layers inside ($\sim 2$D) magnetic flux ribbons.

We proposed a simple relationship between the Mach number of the turbulence and the Mach numbers of shocks that they generate derivedfrom the width of the density PDF.
The relation can be applied to both solenoidally driven or compressively driven turbulence, although the parameterization depends on the nature of the forcing, so it is important again to evaluate the character of the turbulent driving forces.

\acknowledgements

The work of DHP and TWJ was supported in part by the US National Science Foundation through grant AST1211595.
The work of DR was supported by the year of 2014 Research Fund of UNIST (1.140035.01).
This work was carried out in part using computing resources at the University of Minnesota Supercomputing Institute. We express our gratitude to an anonymous referee for helpful comments that improved the presentation of the paper.

\appendix

\section{Enstrophy Generation Across Shock Jumps}
\label{shockvort}


Vorticity and enstrophy generation in shocked flows are fundamental to an understanding of compressible turbulence. To support discussions above we summarize here two simple, analytic ways to establish vorticity and enstrophy generation across shock jumps. We initially ignore magnetic field influences, but will subsequently address those. The first, mathematically more direct approach begins with a form of Euler's equation,
\begin{equation}
\frac{D\vec u}{Dt} = \frac{\partial \vec u}{\partial t} +(\vec u\cdot\nabla)\vec u = -\frac{1}{\rho}\nabla P_g,
\label{euler}
\end{equation}
 expressed explicitly in terms of vorticity using the vector identity, $\vec u \cdot\nabla\vec u = (1/2)\nabla u^2 - \vec u\times\vec\omega$; namely,
\begin{equation}
\vec u \times\vec\omega = \frac{1}{2}\nabla u^2 + \frac{1}{\rho}\nabla P_g + \frac{\partial\vec u}{\partial t}.
\label{Eulera}
\end{equation}
Near a shock with local normal direction $\hat n = \hat q \times \hat z$,  defined by two orthogonal shock tangent directions, $\hat q$ and $\hat z$, equation (\ref{Eulera}) can be projected along $\hat q$ to give,
\begin{equation}
u_n \omega_z = u_z\omega_n -\frac{\partial}{\partial s}\left(\frac{1}{2}u^2\right) - \frac{1}{\rho}\frac{\partial P_g}{\partial s} -\frac{ \partial u_q}{\partial t} ,
\label{uomegaz}
\end{equation}
with $\partial / \partial s = \hat q \cdot \nabla$ measuring variations in this direction along the shock surface. The analogous projection of equation (\ref{Eulera}) in the orthogonal tangent direction, $\hat z$, looks the same, with $\omega_q$ and $u_q$ replacing $\omega_z$ and $u_z$, while the three RHS derivatives (now with $(\partial /\partial s) = \hat z \cdot \nabla$) change sign.

Being based on ideal flow, equation (\ref{uomegaz}) does not apply inside shocks, but by using shock jump conditions, it can be applied to quasi-ideal flows across shocks. We emphasize that the flows can be inhomogeneous and unsteady, so long as the local shock jumps are well defined in space and time. In particular
mass and momentum conservation in hydrodynamical shocks give
\begin{eqnarray}\label{ujumps}
u_{n,2} = \frac{1}{r} u_{n,1},\nonumber\\
u_{q,2} = u_{q,1},\\
u_{z,2} = u_{z,1},\nonumber
\end{eqnarray}
so
\begin{equation}
u_2^2 = u_1^2 - \frac{r^2 - 1}{r^2} u_{n,1}^2,
\label{delusq}
\end{equation}
and
\begin{equation}
P_{g,2} = P_{g,1} + \frac{r - 1}{r}\rho_1 u_{n,1}^2,
\label{delp}
\end{equation}
where the subscripts $1,2$ correspond to upstream and downstream states measured in the local shock rest frame, and $r = 1+\mu = \rho_2/\rho_1$. It is straightforward using equation (\ref{ujumps})
to show that $\delta\omega_n = \omega_{n,2}-\omega_{n,1} = 0$. In addition equations (\ref{ujumps}) guarantee that $\partial (u_{q,2}-u_{q,1})/\partial t = 0$. Using equations (\ref{ujumps}), (\ref{delusq}) and (\ref{delp}) we then obtain the jump in $\omega_z$ as
\begin{equation}
\delta\omega_z = \omega_{z,2} - \omega_{z,1} = \mu\omega_{z,1} -\mu u_{n,1}\frac{1}{\rho_1}\frac{\partial\rho_1}{\partial s}\left(\frac{1}{1+\mu}-\frac{c_{s,1}^2}{u_{n,1}^2}\right)
+\frac{\mu^2}{1+\mu}\frac{\partial u_{n,1}}{\partial s},
\label{delomega}
\end{equation}
where the sound speed, $c_s$, is used in the second term inside the parenthesis, according to barotropic relation $\partial P_g/\partial s = c_s^2 \partial \rho/\partial s$, in order to simplify the form.  Alternate forms of that term, no longer including parenthesis and no longer assuming anything about the equation of state, would be $[\mu/(u_{n,1}\rho_1)]\partial P_1/\partial s = - (\mu/u_{n,1}) Du_{q,1}/Dt$. There is an obvious, analogous expression for $\delta\omega_q$. Equation (\ref{delomega})
is equivalent to equation (4) in \cite{kev09}, who present the result in somewhat different form. Note that we did not assume a steady shock boundary, nor a planar shock, nor did we enforce energy conservation across the transition. Terms including $\partial\mu/\partial s$ that are apparent from insertion of equation (\ref{delusq}) and equation (\ref{delp}) into equation(\ref{uomegaz}) cancel, so equation(\ref{delomega}) also incorporates  compression variation, or equivalently, Mach number variation along the shock surface. As pointed out by \cite{kev09} this result is, then, quite general. 

The first RHS term in equation (\ref{delomega}) represents conservation of circulation across the shock; this enhancement in vorticity components tangent to the shock face just reflects the reduced downstream area of circulation in a plane that includes the shock normal, $\hat n$. The second RHS term including the parentheses represents baroclinic creation of vorticity across the shock when the upstream conditions are inhomogeneous. In an isothermal fluid, such as in our study, $1+\mu = u_{n,1}^2/c_s^2$, so that this term vanishes. 
The third RHS term in equation (\ref{delomega}) accounts for variations in the shock normal speed along its surface; it represents variations in the refraction of streamlines crossing the shock. The variations include nonuniform upstream flows in plane shocks, but also normal velocity variations introduced along shock surfaces due to non-planar shock geometry. In the latter context it is commonly associated with vorticity creation in curved shocks or intersecting shocks \citep{crocco}. For a uniform flow into a shock with radius of curvature, $R$, this term has a magnitude
\begin{equation}
\delta\omega \sim\frac{\mu^2}{1+\mu}\frac{u_1}{R}.
\label{curveshock}
\end{equation}

An alternate derivation of equation (\ref{delomega}) that makes some physical contributions to the vorticity jump more obvious uses terms from $(\vec u\cdot\nabla)\vec u$  directly from equation (\ref{euler}). Again, working with $\omega_z = \partial u_n/\partial q - \partial u_q/\partial n$, we see that we need to evaluate jumps in  two quantities, $\partial u_n/\partial q$ and $\partial u_q/\partial n$. To avoid confusion with notation in the previous form we write here the tangential and normal derivatives as $\partial/\partial q = \nabla\cdot\hat q$ and $\partial/\partial n = \nabla\cdot\hat n$. The first of these is easily determined from the first of the jump conditions (\ref{ujumps}) as,
\begin{equation}
\frac{\partial u_{n,2}}{\partial q} = \frac{\partial}{\partial q} \left(\frac{1}{r} u_{n,1}\right) = \frac{1}{r} \frac{\partial u_{n,1}}{\partial q} - \frac{u_{n,1}}{r^2}\frac{\partial r}{\partial q}.
\label{delun}
\end{equation}
The second can be obtained in a few more steps by projecting equation (\ref{euler}) along $\hat q$ on each side of the shock ($k = 1,2$) to produce, 
\begin{equation}
\frac{\partial u_{q,k}}{\partial n} = -\frac{1}{u_{n,k}}\left[\frac{1}{2}\frac{\partial u_{q,k}^2}{\partial q}+u_{z,k}\frac{\partial u_{q,k}}{\partial z}+\frac{1}{\rho}\frac{\partial P_{g,k}}{\partial q}+\frac{\partial u_{q,k}}{\partial t}\right].
\end{equation}
Now we can, using the second and third jump conditions in equation (\ref{ujumps}), take the difference, $\partial u_{q,2}/\partial n - \partial u_{q,1}/\partial n$ to obtain,
\begin{equation}
\frac{\partial u_{q,2}}{\partial n} = r \frac{\partial u_{q,1}}{\partial n} +\frac{1}{u_{n,1} \rho_1} \left(r \frac{\partial P_{g,1}}{\partial q} -\frac{\partial P_{g,2}}{\partial q}\right).
\end{equation}
Application of the shock pressure jump, equation (\ref{delp}), leads finally to 
\begin{equation}
\frac{\partial u_{q,2}}{\partial n} = r\frac{\partial u_{q,1}}{\partial n} +\frac{r - 1}{u_{n,1}\rho_1}\frac{\partial P_{g,1}}{\partial q} -2\frac{r-1}{r}\frac{\partial u_{n,1}}{\partial q}
-\frac{r-1}{r}\frac{u_{n,1}}{\rho_1}\frac{\partial \rho_1}{\partial q}
-\frac{u_{n,1}}{r^2}\frac{\partial r}{\partial q}.
\label{deluqn}
\end{equation}
The first RHS term in equation (\ref{deluqn}) simply accounts for compression of tangential shear across the shock. The remaining terms incorporate the stresses downstream caused by upstream pressure, density and normal velocity variations tangential to the shock, as well as tangential variations in the shock compression. Note that in combination with equation (\ref{delun}) to create $\omega_z$, this last term is canceled. The others, along with the first term in equation (\ref{delun}), account fully for the vorticity change through the shock. Indeed, the combination $\partial u_{n,2}/\partial q - \partial u_{q,2}/\partial n$ from equations (\ref{delun}) and (\ref{deluqn}) leads once again to $\delta\omega_z$ in equation (\ref{delomega}).

The enstrophy change across the shock, $\delta\epsilon = (1/2) \delta\omega^2$  is simply
\begin{equation}
\delta\epsilon =\delta\vec\omega\cdot\vec\omega_1 + \frac{1}{2}(\delta\omega)^2 = \lbrace \delta\omega_z\cdot\omega_{z,1}+ \delta\omega_q\cdot\omega_{q,1}\rbrace+\frac{1}{2}\left[(\delta\omega_z)^2+ (\delta\omega_{q})^2\right],
\label{delenst}
\end{equation}
where, since $\delta\omega_n = 0$, we have set $ \delta\vec\omega\cdot\vec\omega_1=\delta\omega_z\cdot\omega_{z,1}+  \delta\omega_q\cdot\omega_{q,1}$, and $(\delta\omega)^2 = (\delta\omega_q)^2 +  (\delta\omega_z)^2$.
These can be evaluated from equation (\ref{delomega}) and its $\delta\omega_q$ analogy. All the terms contributing to $\delta\epsilon$ are inherently non-negative, except for the 2nd and 3rd RHS (source) terms in equation (\ref{delomega}) when they appear inside the curly bracket in equation (\ref{delenst}). For example, the contribution of the ``Crocco's'' source term to the enstrophy  jump equation (\ref{delenst}) is $-[\mu^2/(1+\mu) ](\vec\omega\times\nabla)\cdot\vec u_{n,1}$, which can take either sign. Note that even if  $\vec\omega_1 = 0$, the last two RHS terms in equation (\ref{delenst}) will generally be finite and positive.
 
On the whole, except in cases where the incident vorticity magnitude is very small and the shock curvature radius  is very small and negative  (e.g., in some shock intersections), the total change, $\delta\epsilon$, across shocks given by equation (\ref{delenst}) should be positive.
Thus, we expect the integrated quantity $\int\vec u\cdot\nabla\epsilon dn \sim u_{n,1}\delta\epsilon>0$ across most shocks.  That is significant to our analysis of volume integrated enstrophy evolution. It helps explain, in particular, why the volume integral of the $\partial\epsilon/\partial t$ contributor, $F_{comp} = -\nabla\cdot\vec u\epsilon + \vec u\cdot\nabla\epsilon$,  (equations (\ref{eq:enst}), (\ref{eq:enstsource})) provides a good measure of the enstrophy generation by shocks in our turbulent flows (e.g., Figure {\ref{denst}). Additional insights to this relationship come  from the filtered-flow enstrophy analysis in Appendix \ref{filter}.

The presence of magnetic fields can also contribute to and modify vorticity (enstrophy) generation in shocks. Equation (\ref{Eulera}) can be extended to MHD by including Maxwell stress terms,
\begin{equation}
\frac{\vec F_M}{\rho} = -\frac{1}{\rho}\left(\nabla P_B - \vec T\right),
\label{maxwell}
\end{equation}
to the RHS. There are two ways that the Maxwell stresses enter the problem at hand. First, $\vec F_M$ explicitly provides a contribution, $\delta\omega_{B,z}$, to the RHS of
equation (\ref{delomega}),
\begin{equation}
\delta\omega_{B,z} = -\frac{1}{u_{n,1}\rho_1}\left(\frac{\partial P_{B,2}}{\partial s}-(1+\mu)\frac{\partial P_{B,1}}{\partial s}
+ \hat q\cdot\vec T_2 - (1+\mu) \hat q\cdot\vec T_1\right).
\label{delomegab}
\end{equation}
In addition, the jump conditions expressed in  equations (\ref{ujumps}), (\ref{delusq}) and (\ref{delp}), will be modified because of the anisotopic nature of $\vec T$. The latter influences are complex and depend on the nature of the shock (fast or slow mode).  A couple of simple examples, however, are sufficient to establish the character of $\delta \omega_{B,z}$.  First, consider the case of a plane shock interacting with a one-dimensional (1D) magnetic field aligned with the shock normal, $B(x_t)\hat n$. Then the only change from equation (\ref{delomega}) comes from the magnetic pressure, which does not in this case change across the shock, while the gas pressure does change according to equation (\ref{delp}). The total pressure is no longer barotropic even in an isothermal gas, and there is an added term to equation (\ref{delomega}),
\begin{equation}
\delta\omega_{B,z} = 2\mu\frac{P_B}{u_{n,1}\rho_1}\frac{1}{B}\frac{\partial B}{\partial s} = 2\mu\frac{u_{n,1}}{\mathcal{M}_A^2}\frac{1}{B}\frac{\partial B}{\partial s},
\end{equation}
where $\mathcal{M}_A = u_{n,1}/v_A$ is the local Alfv\'enic Mach number of the shock.

For a second example again consider a uniform magnetic field, $B$, now interacting with a curved shock having a radius of curvature, $R$. Then the inclination of the field to the shock will vary with location along the shock, as will the magnetic pressure jump even if $\mu$ is constant. In addition, the magnetic tension forces will modify tangential velocity jumps, adding terms to equation (\ref{delusq}) of order $u_t u_n/\mathcal{M}_A^2$. The net result of the magnetic field can be incorporate into a term with magnitude,
\begin{equation}
\delta\omega_{B,z} \sim \frac{1}{\mathcal{M}_A^2}\frac{u_1}{R}.
\end{equation}

Since these magnetic field-induced vorticity sources scale inversely with Alfv\'enic Mach number squared compared to hydrodynamical sources, they will be relatively small in the context of ICMs \citep[e.g.,][]{ryu08}. On the other hand, these influences would become significant when magnetic fields are dynamically important on scales $\sim R$.

\section{Filtered Flows as a Probe of  Enstrophy Generation in Turbulence Shocks}
\label{filter}

The methods used in Appendix \ref{shockvort} to quantify the generation of vorticity and enstrophy across shocks are based mathematical discontinuity at shocks with well-defined states on both sides of the discontinuity. On the other hand, turbulent flows and the shocks they generate are by nature complex, and the gradient operations involved in evaluating the critical dynamical variables are not, in general, analytically well-defined across shocks. In addition, ICM shocks are effectively collisionless, so that their internal structures are both complex and unsteady. When finite difference derivatives are involved in computing shock transitions these kinds of issues are similarly obvious. Consequently, in simulations estimates of derivative-based  quantities across the numerical shocks might not be reliable, especially near locations of strongly curved or intersecting shocks, where the results of Appendix \ref{shockvort} suggest the most significant shock-associated vorticity generation probably takes place. Similar concerns exist at slip surfaces, where, again spatial derivatives are not necessarily properly defined.

A simple and effective strategy to ameliorate these issues commonly employed in turbulence studies is to work with properly filtered flow variables that eliminate both the mathematical and numerical problems.  In particular, derivatives of the velocity field can be expressed in terms of filtered (smoothed) values together with a well-behaved sub-filter shear or Reynolds stress.  Discontinuities formally disappear, so all the metrics are well represented within the filtered flow. This provides not only a mathematical and numerical framework that is well-founded, but also, as we show, offers useful insights into the manner in which vorticity is generated within shocks. Here we introduce such an analysis and apply it, in particular, to examine the shock generation of vorticity in flows that are initially irrotational, $\vec\omega(t,\vec x) = 0$, and to confirm our conclusion from unfiltered flow analyses of our simulations that the $F_{comp}$ source term defined in equation (\ref{eq:enstsource}) is predominantly associated with shocks and represents the dominant shock-related source of enstrophy in isothermal flows. This finding also reinforces the previous finding that the evolution of vorticity across shocks is predominantly determined by basic conservation of mass and momentum across those shocks.

We start with the unfiltered continuity and momentum equations for a compressible, ideal fluid (for simplicity setting $\vec B = 0$ and $\nu = 0$ and ignoring the forcing term),
\begin{equation} \label{mass_conservation}
\frac {\partial \rho}{\partial t} + \nabla \cdot ( \rho \vec u ) = 0,
\end{equation}
\begin{equation} \label{momentum_conservation}
\frac {\partial \rho \vec u}{\partial t} +  \nabla \cdot ( \rho \vec u \vec u ) = - \nabla P.
\end{equation}

Given any spatial normalized convolution filter
\begin{equation} \label{raw_filter}
\bar Q(\vec x) = \int g(\vec x - \vec x_1)  Q(\vec x_1) d^3 \vec x_1
\end{equation}
the corresponding Favre (or mass weighted) filter \citep{favre83} is
\begin{equation} \label{favre_filter}
\tilde Q = \overline {\rho Q} / \bar \rho.
\end{equation}
The resulting equations for evolution of the filtered set of fields $(\bar \rho, \bar P, \tilde{\vec u})$
become
\begin{equation} \label{mass_favre}
\frac {\partial \bar \rho}{\partial t} +  \nabla \cdot ( \bar \rho \tilde{\vec u} ) = 0,
\end{equation}
\begin{equation} \label{momentum_favre}
\frac {\partial \bar \rho \tilde{\vec u}}{\partial t} +  \nabla \cdot ( \bar \rho \tilde{\vec u} \tilde{\vec u})
= -  \nabla \bar P -  \nabla \cdot \vec{\vec \tau},
\end{equation}
where $\vec{\vec \tau}$ is the sub-filter scale Reynolds stress tensor,
\begin{equation} \label{sfs_stress}
\vec{\vec \tau} = \overline{\rho \vec u \vec u} - \bar \rho \tilde{\vec u} \tilde{\vec u}.
\end{equation}

The resulting equation for the Favre-filtered velocity  is
\begin{equation} \label{velocity_favre}
\frac {\partial \tilde{\vec u}}{\partial t} + (\tilde{\vec u} \cdot \vec \nabla) \tilde{\vec u}
= - \frac{\nabla \bar P}{\bar \rho} - \frac{\vec \nabla \cdot \vec{\vec \tau}}{\bar \rho}
\end{equation}

The curl of equation (\ref{velocity_favre}) provides an equation for the evolution of the vorticity of the filtered velocity,  $\tilde {\vec \omega} = \nabla\times \tilde{\vec u}$,
\begin{equation} \label{vorticity_favre}
\frac {\partial \tilde{\vec \omega}}{\partial t} + (\tilde{\vec u} \cdot \vec \nabla) \tilde{\vec \omega}
= (\tilde{\vec \omega} \cdot \vec \nabla) \tilde{\vec u}
- \tilde{\vec \omega} (\vec \nabla \cdot \tilde{\vec u})
+ \frac{\vec \nabla \bar \rho \times \vec \nabla \bar p}{{\bar \rho}^2}
- \vec \nabla \times \left(\frac{ \vec \nabla \cdot \vec {\vec \tau}}{\bar \rho} \right).
\end{equation}
The concerns about spatial derivatives outlined at the start of this appendix are avoided with this equation, since  $\tilde {\vec \omega}$ represents the curl of the filtered velocity, not a filtering of $\nabla\times\vec u$.

By taking the inner product of  $\tilde {\vec \omega}$ with equation \ref{vorticity_favre} and rearranging some terms
we get an equation for evolution of the enstrophy density $\epsilon_f = \frac{1}{2} {\tilde \omega}^2$ of the Favre-filtered
 flow that is very similar in form to equation (\ref{eq:enst}) for the enstrophy of the unfiltered flow; namely,
\begin{equation} \label{enstrophy_source_favre}
\frac {\partial \epsilon_f}{\partial t}=\tilde F_{adv} + \tilde F_{stretch} + \tilde F_{comp} + \tilde F_{baroc} + \tilde F_{sfs} = \tilde F_{adv} + \tilde F_{tot},
\end{equation}
where now
\begin{eqnarray}\label{eq:favresource}
\tilde F_{adv} =- \nabla\cdot (\tilde{\vec u} \epsilon_f) =-( \epsilon_f\nabla\cdot\tilde{\vec u} + \tilde{\vec u}\cdot\nabla\epsilon_f),\nonumber\\
\tilde F_{stretch} = \tilde{\vec\omega}\cdot (\tilde{\vec\omega}\cdot\nabla)\tilde{\vec u} =  2\epsilon_f (\tilde{\hat\omega}\cdot\nabla)\tilde{\vec u} \cdot \tilde{\hat\omega},\nonumber\\
\tilde F_{comp} = -\epsilon_f \nabla\cdot \tilde{\vec u} = \frac{\epsilon_f}{\bar\rho}\frac{d\bar\rho}{dt} = -\nabla\cdot (\tilde{\vec u} \epsilon_f) + \tilde{\vec u}\cdot\nabla\epsilon_f,\\
\tilde F_{baroc} = \frac{\tilde{\vec\omega}}{\bar\rho^2} \cdot (\nabla\bar\rho\times\nabla\bar P),\nonumber\\
\tilde F_{sfs} = - \tilde{\vec \omega} \cdot \left[ \vec \nabla \times \left(\frac{ \nabla \cdot \vec {\vec \tau}}{\bar \rho} \right) \right],\nonumber
\end{eqnarray}
are flux and source terms analogous to those in equation (\ref{eq:enstsource}) expressed in terms of filtered quantities,  and
\begin{equation}\label{eq:ftot}
\tilde F_{tot} = \tilde F_{stretch} + \tilde F_{comp} + \tilde F_{baroc} + \tilde F_{sfs},
\end{equation}
while remembering in our isothermal simulations $\tilde F_{baroc} = 0$. In this derivation we have neglected magnetic and viscous stresses, so terms $\tilde F_{mag}$ and $\tilde F_{diss}$ are omitted, but a new, baroclinic-like term has appeared, $\tilde F_{sfs}$, which represents enstrophy generation through anisotropic, sub-filter scale Reynolds stresses. In what follows we emphasize again that $\tilde{\vec\omega}$ and $\epsilon_f$ 
represent solenoidal properties of the filtered velocity field, not a filtering of the curl of the raw velocity field. 

Since the volume integral of $\tilde F_{adv}$ vanishes in our periodic domain, we focus on the three net-contributor source terms  in equation (\ref{eq:favresource}), designating their sum as $\tilde F_{tot}$ (equation (\ref{eq:ftot}) with $\tilde F_{baroc} = 0$).   For a consistency check we have verified numerically from our simulation data that, within about 10\%, $\partial\epsilon_f(\vec x)/\partial t - \tilde F_{adv}(\vec x) \approx \tilde F_{tot}(\vec x)$ over the grid, with  $\partial\epsilon_f(\vec x)/\partial t$ estimated explicitly from differences in $\epsilon_f(\vec x)$ constructed from  data in the closest saved  time steps. This means that  enstrophy accounting including only ideal terms provides a pretty good match to the data.

As a simple, easily constructed filter, $g(r)$, we applied a spatial convolution based on iterating a top-hat filter.
In particular, the top-hat filter averaged a 3x3x3 block of cells to produce the filtered value in the central
cell.  This top-hat filter was then iterated several times, which is an efficient way to implement a convolution with the Gaussian form,
\begin{equation} \label{gaussian_filter}
 g(r) = {1 \over { ( \sigma \sqrt {2 \pi} )^3 } } e^{-r^2/(2 \sigma^2)},
\end{equation}
where 
$\sigma = 0.826 * \sqrt N$.
For $N > 3$ this is quite accurate, and the effective convolution function is impressively isotropic.
In our analysis we used $N = 8$, which is equivalent to a Gaussian convolution with $\sigma = 2.336 \Delta x$, where $\Delta x$ is the size of one computational cell.  At the price of a modest smoothing in the flows we thus obtained a substantial gain in ability to extract consistent solenoidal measures of the simulated flows in regions of sharp spatial gradients that would, otherwise, be difficult to characterize.

Our particular interest in this Appendix is vorticity/enstrophy generation in complex shock structures. So, from here we focus on the application of this formalism to the early evolution of enstrophy in our compressively forced, $f_s = 0$, simulations, where complex, intersecting shocks develop directly from the forced compressions into an initially irrotational flow. We examined  in our simulation data both the global and local behaviors of the filtered-flow quantities in equations (\ref{enstrophy_source_favre}) and (\ref{eq:favresource}).  With regards to the former we found, with the exception of some very early contributions from $\tilde F_{sfs}$ in the filtered flow, that the time evolution of the volume integrated filtered-flow enstrophy flux and source terms in equation (\ref{eq:favresource}) agreed well with the analogous unfiltered-flow flux terms in equation (\ref{eq:enstsource}) that are shown in Figure \ref{denst}. 

In reference to Figure \ref{denst} and as noted before, the net early growth of the unfiltered enstrophy was due almost entirely to $F_{comp}$ once the first shocks formed ($t\sim 2$) and until solenoidal motions became broadly distributed ($t \sim 15-20$), when $F_{stretch}$ became competitive.  On the other hand, $F_{comp}=-\epsilon\nabla\cdot\vec u\ne 0$ only if $\epsilon\ne 0$, so some seed enstrophy is needed for that term to contribute. In our isothermal flows the baroclinic source term, $F_{baroc}$, vanishes; at early times our magnetic fields are very weak, so $|F_{mag}|$ is quite small. In the  earliest unfiltered flow, before any shocks developed, mostly numerical truncation, noise was available to seed very small amounts of vorticity. On the other hand,  once shocks formed the necessary seeds could come from anisotropic flow stresses that are  represented by the $\nu\nabla\times\vec G$ term in viscous flows, or by the analogous Reynolds stress term, $\tilde F_{sfs}$, in the ideal, filtered flow in the present discussion. Indeed, we found in the interval $2\la t \la5$ of our $f_s = 0$ simulations that the net contributions to enstrophy growth in the filtered flow came in comparable amounts from $\tilde F_{sfs}$ and $\tilde F_{comp}$. After $t \sim 5$ $\tilde F_{sfs}$ dropped rapidly to near zero, leaving $\tilde F_{comp}$ alone as the dominant contributor to $\partial\epsilon_f/\partial t$ before $t \sim 20$, just as for the unfiltered flow. 

Close examination showed, in addition, that the enstrophy production in the filtered flow during the early time period was highly episodic and strongly correlated spatially with shocks, and with shock intersections, especially. That behavior is obvious in Figure \ref{compfig}, taken at $t = 8$ from an $f_s = 0$ simulation and displaying enstrophy sources in a 2D slice  containing a complex of intersecting shocks penetrating a region of irrotational flow. Figure \ref{scut} represents a horizontal (left-to-right) line cut  across the middle of this complex at the same time, showing the total enstrophy source term, $\tilde F_{tot}$, as well as the dominant contributing terms, $\tilde F_{sfs}$ and $\tilde F_{comp}$.  The sharp negative spike in $\tilde F_{tot}$ near x = 0 (also visible in Figure \ref{compfig}) comes from $\tilde F_{stretch}$ (not plotted), revealing the generation of shear within the forward shocks that is enhanced by stretching across the rear, oblique shock along this line. The region to the  left is previously unshocked, so irrotational. Keeping in mind that the quantities shown are derived from Gaussian-filtered (broadened) flow fields, it is then evident that as the flow enters the shock complex it is the sub-filter Reynolds stress term, $\tilde F_{sfs}$, that creates the seed enstrophy, but that as the flow proceeds through the shock complex the compressive term, $\tilde F_{comp}$, becomes a comparable player, while the stretching term, $\tilde F_{stretch}$, can contribute downstream.

\section{Determining Shock Strength Distributions in Grid Simulations}
\label{shockfinder}


Physical shocks are generally approximated mathematically as 2D surfaces; that is, the shock transition becomes vanishingly thin. In grid-based numerical simulations, on the other hand, a shock transition is usually distributed across several grid cells, making accurate identification of the shock surface, as well as a proper determination of the shock normal and shock transition properties challenging. Several approaches to this problem have been used before \citep[e.g.,][]{ryu03,vazza11}. Here we outline a new post-processing algorithm that can be used to construct and analyze shock surfaces in simulations such as ours that lead to complex shock distributions.  We find it to be especially flexible and also quite robust in identifying and cataloging weak shocks. We also find that it gives results very consistent with an updated version of the method in \cite{ryu03}. Our immediate goal is to compute the area weighted probability distribution of shock Mach numbers. Other applications of the method, including those that especially utilize the information it extracts about the shape of shock surfaces, are obvious. 

We assume availability of a 3D snapshot of a simulated flow, and in particular the spatial distribution in cells of the velocity and density fields, $\vec u$ and $\rho$. In brief, the procedure consists of partitioning the computational volume into small cubes, sized so that one can fully enclose a numerical shock transition, but  is unlikely to enclose multiple shocks. Cubes actually containing shocks are identified, the location of the shock surface within the cube, its orientation and strength are determined. Those surfaces can then be displayed and/or their properties analyzed and tabulated.

The initial step is a division of the computational volume into small cubes containing $n_c\times n_c\times n_c = N_c$ cells. The cubes should be large enough to span numerical shock transitions, but we also want to assume at most one shock structure exists inside a given cube. In those cases where shocks intersect within a cube we make the simplifying assumption that the structure can be interpreted in terms of a single shock surface. We have found $n_c = 5$ (so $N_c = 125$) to be a good practical choice.

Two criteria are used to determine which cubes contain a shock surface. First, there must be a sufficiently strong compression rate ($(1/\rho) d \rho/dt = - \nabla\cdot\vec u > 0$) in some fraction of the cube, and second, variations in $\rho$ must be primarily in one direction. The compression rate criterion includes two steps. We require at least a critical minimum number of cells in a cube, $N_{s,c}$, with $\nabla\cdot\vec u < 0$ and that the average divergence  in those $N_s$ cells is less than a critical value, $du_{crit}$. Specifically, inside each cube we compute
\begin{equation}
N_s = \sum_{N_c}  h(-\nabla\cdot\vec u),
\end{equation}
\begin{equation} \label{step}
h(q) = \left\{
\begin{matrix}
0, q \leq 0 \\
1, q > 0
\end{matrix}
\right\},
\end{equation}
and compare $N_s$ to $N_{s,c}$. We have found $N_{s.c} = 8$ to be a good practical choice.
The average compression rate over these cells is
\begin{equation} \label{avc}
<\nabla \cdot \vec u>_s = \frac {1}{N_s} \sum {h(-\nabla \cdot \vec u) \nabla \cdot \vec u},
\end{equation}
which is compared to $du_{crit}$. We have found $du_{crit} = -1\times (\nabla\cdot\vec u)_{max}$, where $ (\nabla\cdot\vec u)_{max}$ measures the maximum rarefaction rate in the simulation box, to be a reliable and effective choice in turbulent flows.

The local variation in the density, $\rho$, is characterized by its gradient.  A quantitative measure of the mean square variations
of $\rho$ in different directions averaged over each cube can be constructed in terms of the average of the tensor product of
$\nabla \rho$ with itself; namely,
\begin{equation} \label{amatrix}
\vec {\vec a} = < \vec \nabla \rho  \vec \nabla \rho >_{cube} = \sum_1^3 \lambda_i \hat e_i.
\end{equation}
This is a real and symmetric matrix that has positive, real eigenvalues,
 $\lambda_1 \geq \lambda_2 \geq \lambda_3 \geq 0$,
with $\lambda_1+\lambda_2+\lambda_3 =  <|\nabla \rho|^2>_{cube}$, and with orthogonal, unit eigenvectors, $\hat e_1, \hat e_2, \hat e_3$, related to $\vec {\vec a}$ as above.
The three eigenvalues correspond to the mean square variation of $\rho$ in the direction of the
corresponding eigenvector.  If $\hat e_1$ is set to be the principle direction
of variation, $\lambda_1$ is the largest eigenvalue.  To the extent that the faction of mean square variation along $\hat e_1$
\begin{equation} \label{frace1}
f_1 = \frac {\lambda_1} {\lambda_1 + \lambda_2 + \lambda_3}
\end{equation}
is close to unity, variation in the other, orthogonal directions is negligible, and the variation of $\rho$ is
primarily along the direction of $\hat e_1$.  Our condition for variations in $\rho$ is
\begin{equation} \label{frac1}
f_1 > f_{crit},
\end{equation}
where $f_{crit}$ is less than but close to unity. We have found $f_{crit} = 0.8$ to be an effective choice.

Having used the above steps to establish the presence and orientation of a shock inside a cube,  we next measure 1D density and velocity profiles along the shock normal, $\hat e_1$, determine the location of the shock surface inside the cube and calculate the shock jump conditions, including, for instance, shock Mach number (given an equation of state -- isothermal in the present case). 
We construct the requisite density and velocity profiles by
binning their cell values against the coordinate $s = \hat e_1 \cdot \vec x$, where
$\vec x$ is the computational  grid coordinate.  We use grid values inside a cylinder centered on the cube with a
cylindrical axis aligned along $\hat e_1$.  The radius of the cylinder can be varied, but we find that half the size of the cube works well, since this is small enough to have little variation in directions transverse
to $\hat e_1$, but large enough to include a few dozen cells in the averages in the $s$ bins.
The length of the cylinder along $\hat e_1$ extends just beyond the cubical volume in order to capture the variation of density and velocity near the the edge of the cube.  We then look for a well defined shock jump; i.e., rapid changes in the density and velocity across a few cells with
relatively constant values on either side of this transition.  Finally, we assign values on either side of the jump as the pre- and
post-shock values of $\rho$ and $\vec u$, which are used to establish the shock jump.  If no well-defined shock jump is found, the cube is rejected as a shock container.  If the cube is accepted, we use the jump centroid  to locate the shock transition, which, with $\hat e_1$, locates and orients the shock plane within the cube.

Once we have located the shock plane we can compute its effective area inside a cube. To do this we first locate the $N_i$ intersection points of the shock plane with the cube edges, $\vec E_i$. In a given shocked cube this will range between $N_i = 3$ and $N_i = 6$. A central point in the plane within the cube can be generated from a linear, vector average of all these cube edge  intersections. Those points can be sorted into a list that runs, for example, clockwise around the central point of the plane. Then a sequence of point pairs can be combined with the central point to construct $N_i$ triangles, whose areas can be added to determine the shock area inside the cube. Appropriate shock properties, such as Mach number can be assigned to the area.  Additional steps to be discussed elsewhere can be used to establish other geometric properties of the shock surface, such as its curvature. 


\clearpage


\begin{deluxetable} {cccccccc}
\tablecaption{Model Summaries}
\tablewidth{0pt}
\tablehead{
\colhead {Model} & \colhead{$f_s$} & \colhead{Grid ($N^3$)} 
& \colhead{$t_{end}$} & \colhead{$R_{eH}$} & \colhead{$L_{I,u}/L_0$} & \colhead{$t_{e, I}$} & \colhead{$\langle P_{B,end}\rangle$}
}
\startdata
S1K    & 1.0 & $1024^3$   &160 & 556 &0.447 & 8.7 & $4.4\times 10^{-2}$\\
S2K  & 1.0 &  $2048^3$  & 130 & 1392 & 0.445 & 9.0 & $5.5\times 10^{-2}$\\
ChK & 0.0 & $512^3$   &  625 & 242 & 0.480 & 12.3 & $1.2\times 10^{-3}$ \\
C1K & 0.0 & $1024^3$   & 120 & 618 & 0.484 & 11.5 & $7.4\times 10^{-4}$\\
\enddata
\tablecomments{All simulations were done in a periodic box of size, $L_0 = 10$ in code units, using an isothermal equation of state, $P_g \propto \rho$, mean density, $\langle\rho\rangle = \rho_0 =  1$, mean gas pressure, $\langle P_g\rangle = P_{g,0} = 1$, so sound speed, $c_s = 1$. Time units are, then, $0.1 L_0/c_s$. The initial magnetic pressure, $P_{B,0} = P_{g,0}/\beta_0$ with $\beta_0 = 10^6$, so Alfv\'en speed  $v_A = \sqrt{2P_B/\rho} = 1.4 \times 10^{-3}$. The parameter $f_s$ measures the fraction of forcing power in solenoidal motions. The peak driving scale, $L_d \approx (2/3) L_0$ (see \S 2). The hydrodynamical integral scale, $L_{I,u}$, is defined in equation \ref{eq:iscale}, while the eddy turnover time at $L_{I,u}$, $t_{e, I} = L_{I,u}/u_{RMS} = (L_{I,u}/L_d) t_d$, where $t_d = L_d/u_{RMS}$ is the ``driving time''. The effective hydrodynamic Reynolds number, $R_{e,H} = (L_{I,u}/\ell_d)^{4/3}$ with $\ell_d = 4\Delta x \equiv 4L_0/N$.} Plasma $\beta$ values at $t_{end}$ are $\langle\beta_{end}\rangle = 1/\langle P_{B,end}\rangle$.
\label{table1}
\end{deluxetable}

\clearpage


   \begin{figure}
   \centering
   \includegraphics[width = 0.75\textwidth]{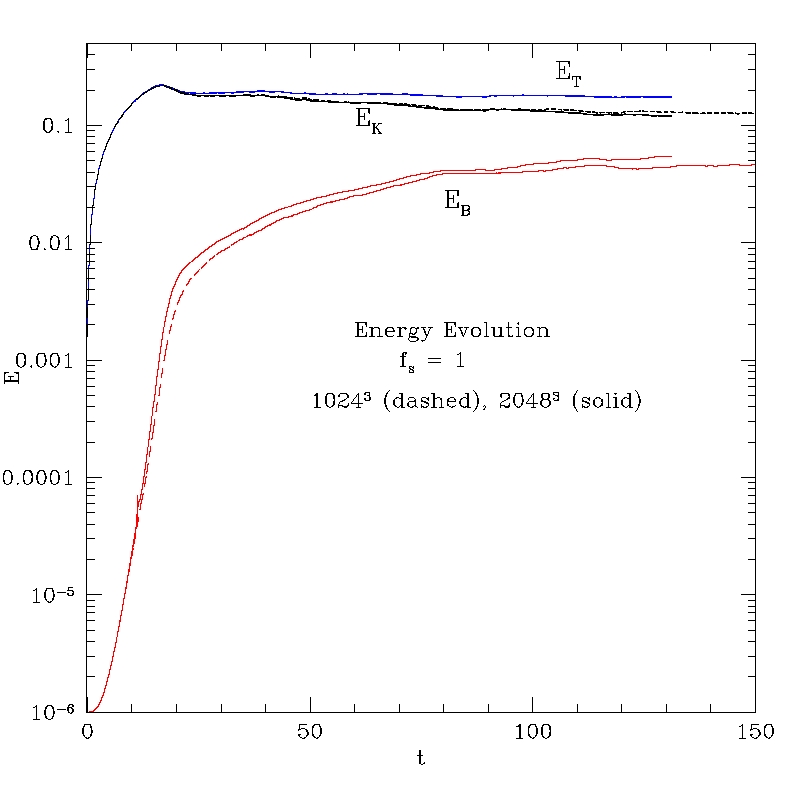}
   \caption{Evolution of kinetic, $E_K$,  magnetic, $E_B$ and total, $E_T$,
   energies for the S1K and S2K simulations of isothermal, compressible, solenoidally driven ($f_s = 1$) MHD turbulence.}
              \label{sol:evol}%
    \end{figure}

\clearpage

\begin{figure}
   \centering
   \includegraphics[width = 0.75\textwidth]{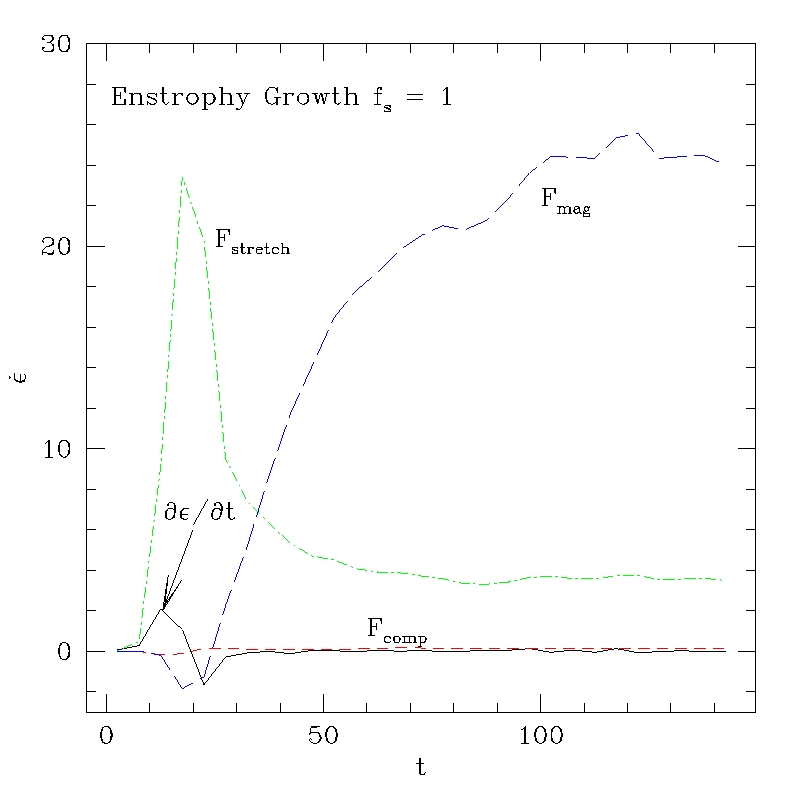}
   \caption{Enstrophy growth rate, $\partial\epsilon/\partial t$, and the contributing source terms for the S1K simulation with $f_s = 1$. Quantities are dimension $t^{-3}$  in code time  units.}
              \label{soldenst}%
    \end{figure}

\clearpage

   \begin{figure}
   \centering
   \includegraphics[width = 0.75\textwidth]{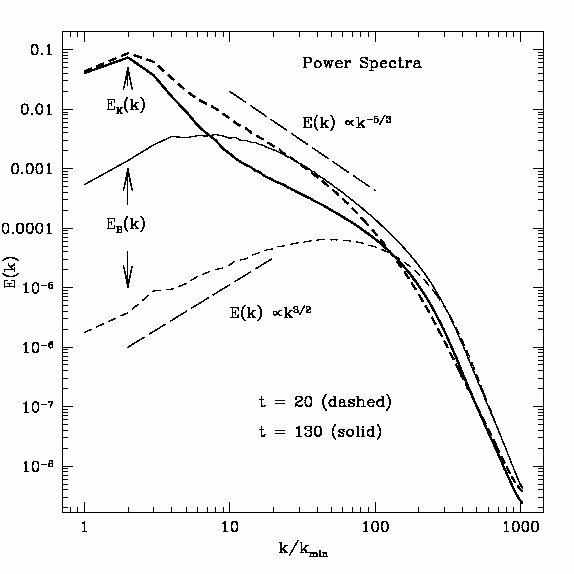}
   \caption{Power spectra, $E(k)$,
   of kinetic ($E_K(k)$)  and magnetic ($E_B(k)$) energies at $t = 20$
   and $t=130$ in the S2K simulation with $f_s = 1$. The long-dash line segment represents a Kolmogorov slope.}
              \label{sol:spec}%
    \end{figure}

\clearpage

\begin{figure}
    \centering
      \includegraphics[width=0.48\textwidth]{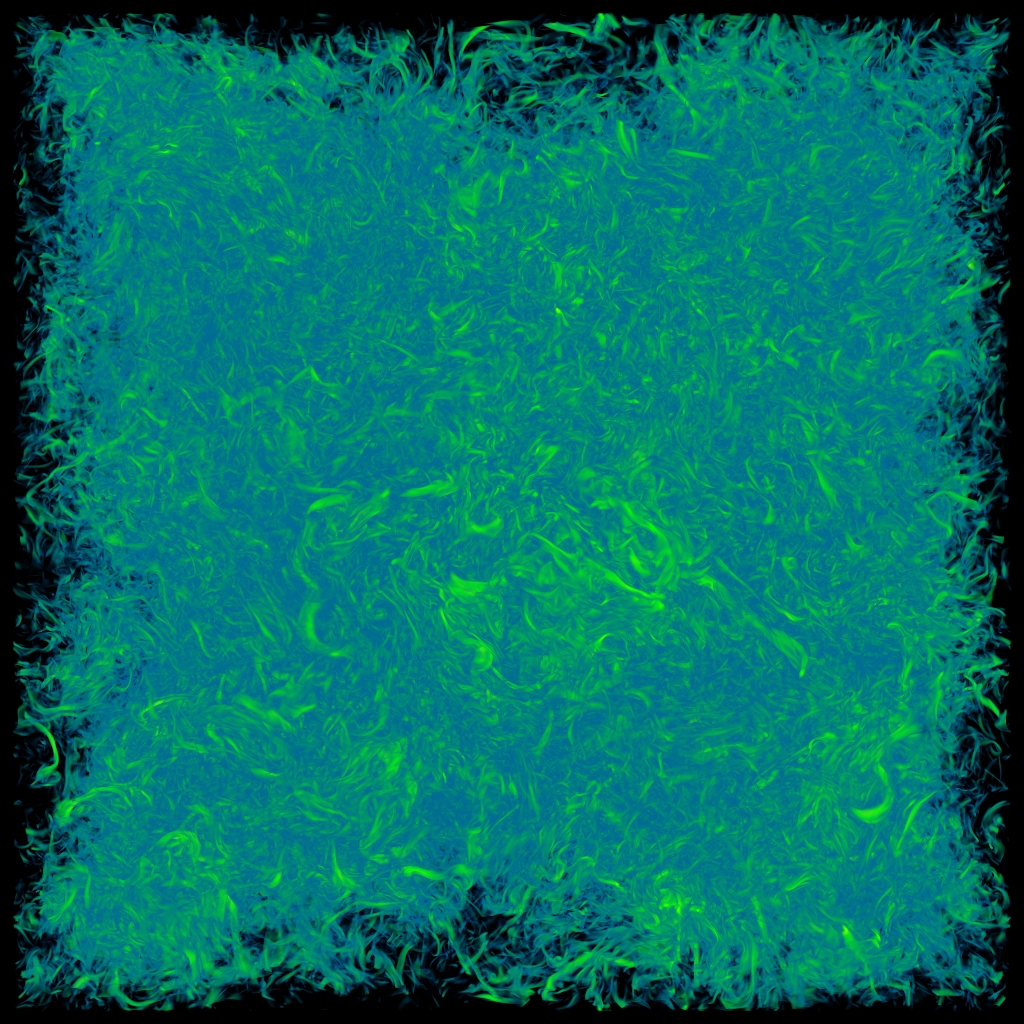}
      \includegraphics[width=0.48\textwidth]{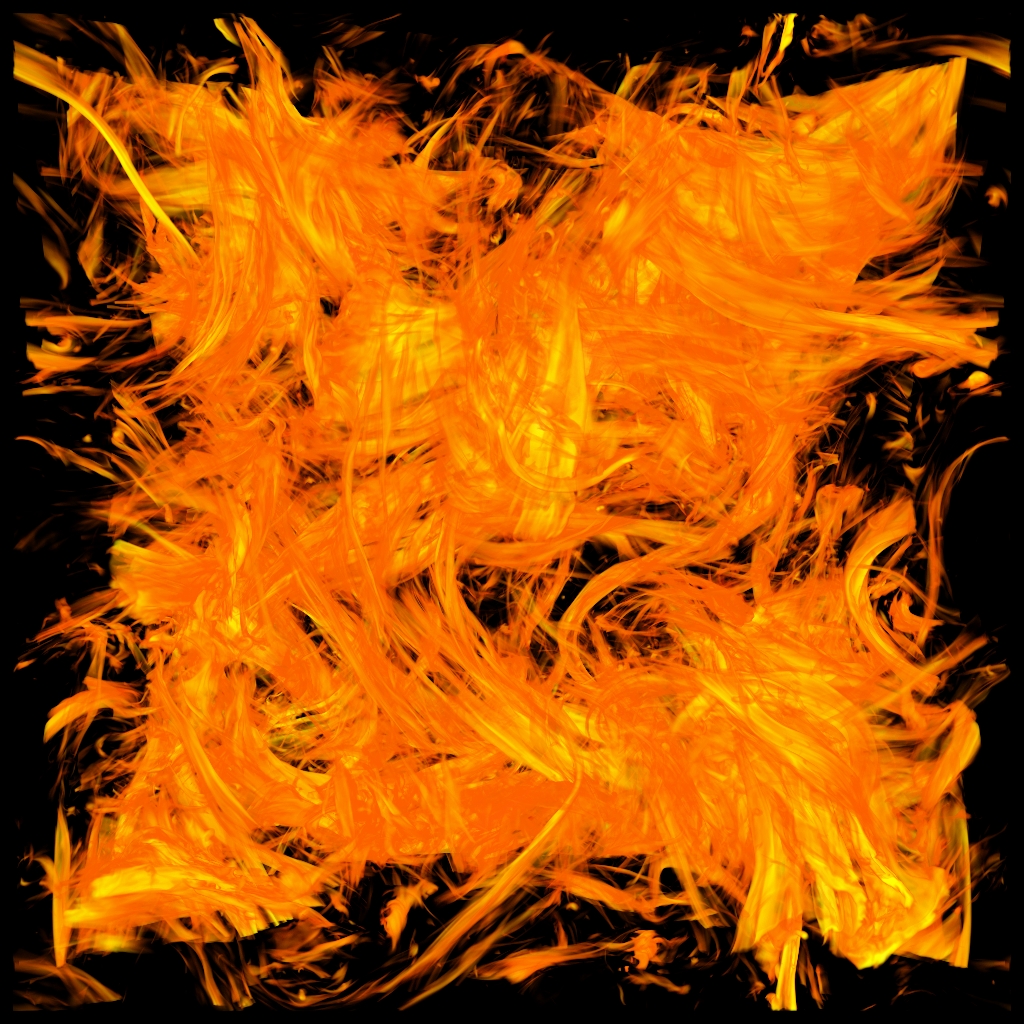}
    \caption{Magnetic energy density, $E_B$,  volume renderings in the SK1 simulation at $t  =20$  (Left) and $t = 130$ (Right). ``Cool'' is weak; ``hot'' is strong. Opacities are chosen to isolate stronger fields.}
    \label{solimages}
\end{figure}

\clearpage

   \begin{figure}
   \centering
   \includegraphics[width = 0.75\textwidth]{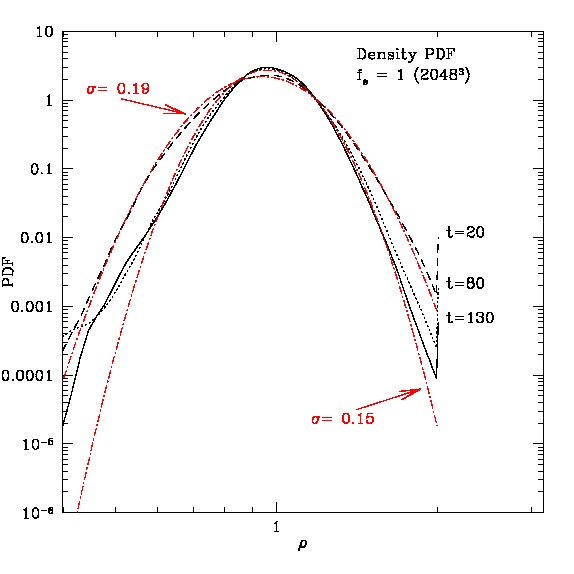}
   \caption{PDF of density in the S1K simulation at three times. Red dot-dash curves represent lognormal distribution fits at $t = 20$ and $t = 130$}
              \label{sol:pdf}%
    \end{figure}

\clearpage

 \begin{figure}
   \centering
   \includegraphics[width = 0.8\textwidth]{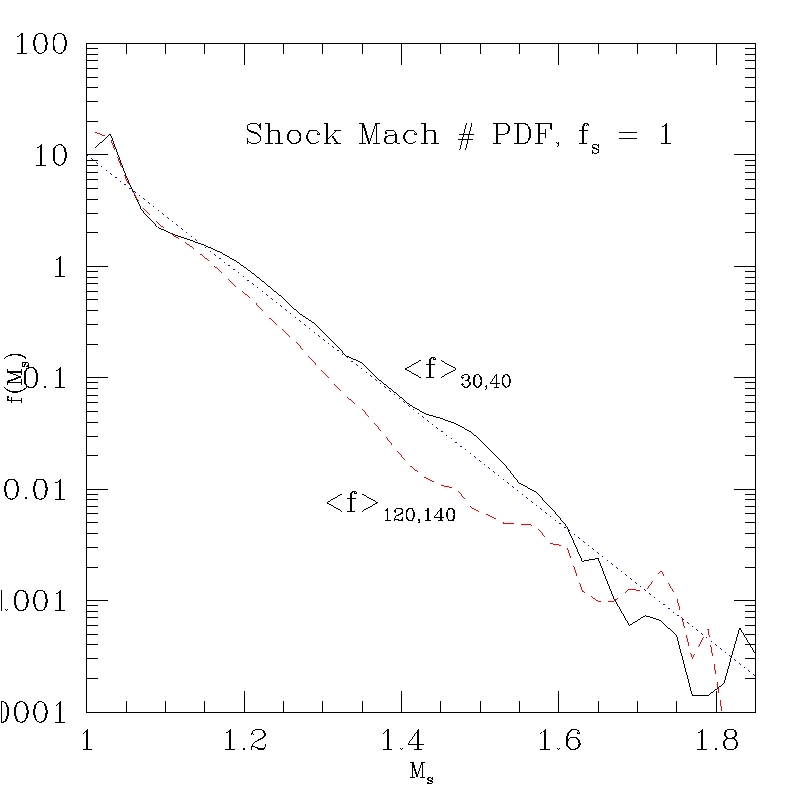}
   \caption{PDF of shock Mach numbers in the S1K simulation. Each of the data curves $\langle f\rangle_{30,40}$ and $\langle f\rangle_{120,140}$ corresponds to an average for two indicated times. The dotted line represents an exponential PDF in the variable $\mathcal{M}-1$ with a characteristic $\mathcal{M}_{sc} =  1.08$.}
              \label{solmachdist}%
    \end{figure}

\clearpage

   \begin{figure}
   \centering
   \includegraphics[width = 0.75\textwidth]{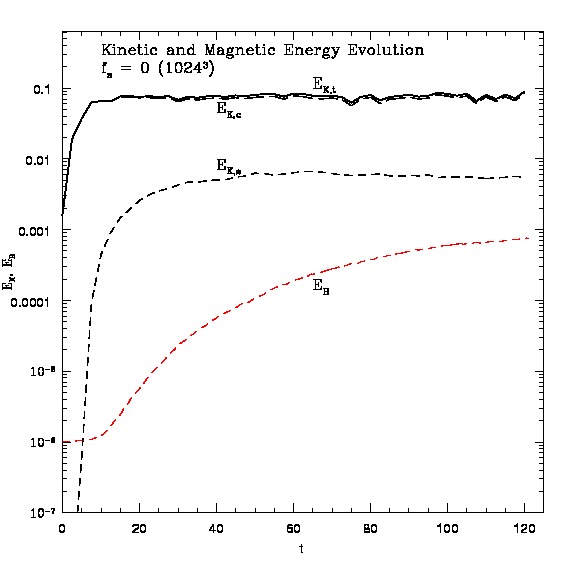}
   \caption{Evolution of kinetic energy, $E_{K,c}$ and $E_{K,s}$, and magnetic energy, $E_B$, for the C1K simulation of compressively driven ($f_s = 0$) MHD turbulence.}
              \label{comp:evol}%
    \end{figure}

\clearpage

 \begin{figure}
   \centering
   \includegraphics[width = 0.8\textwidth]{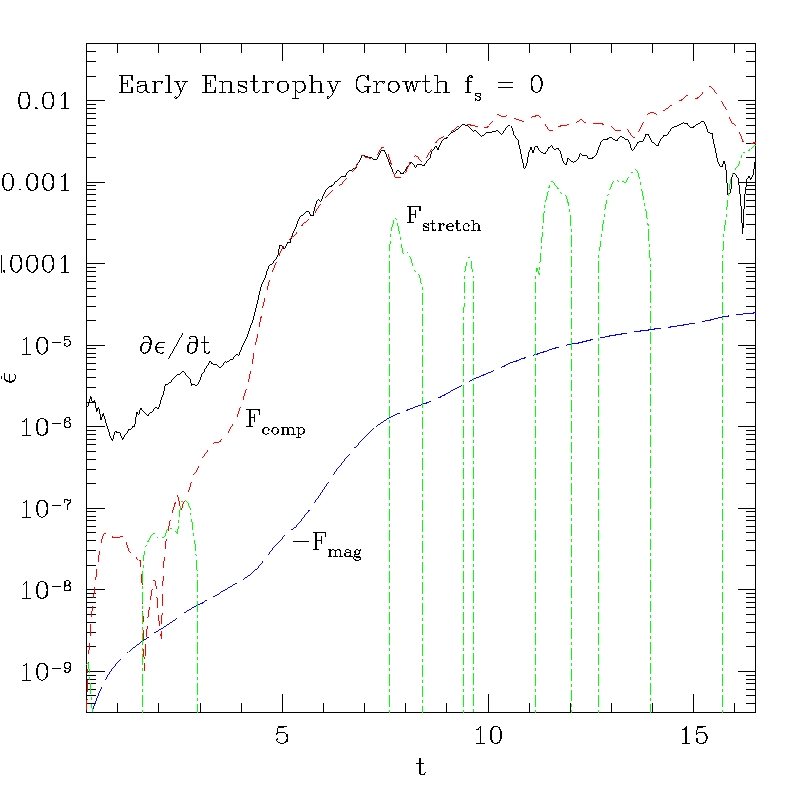}
   \caption{Early evolution of the enstrophy growth rate, $\partial\epsilon/\partial t$, along with
   contributing terms from equation (\ref{eq:enst}) for the C1K simulation.}
              \label{denst}%
    \end{figure}

\clearpage

\begin{figure}
    \centering
      \includegraphics[width=0.75\textwidth]{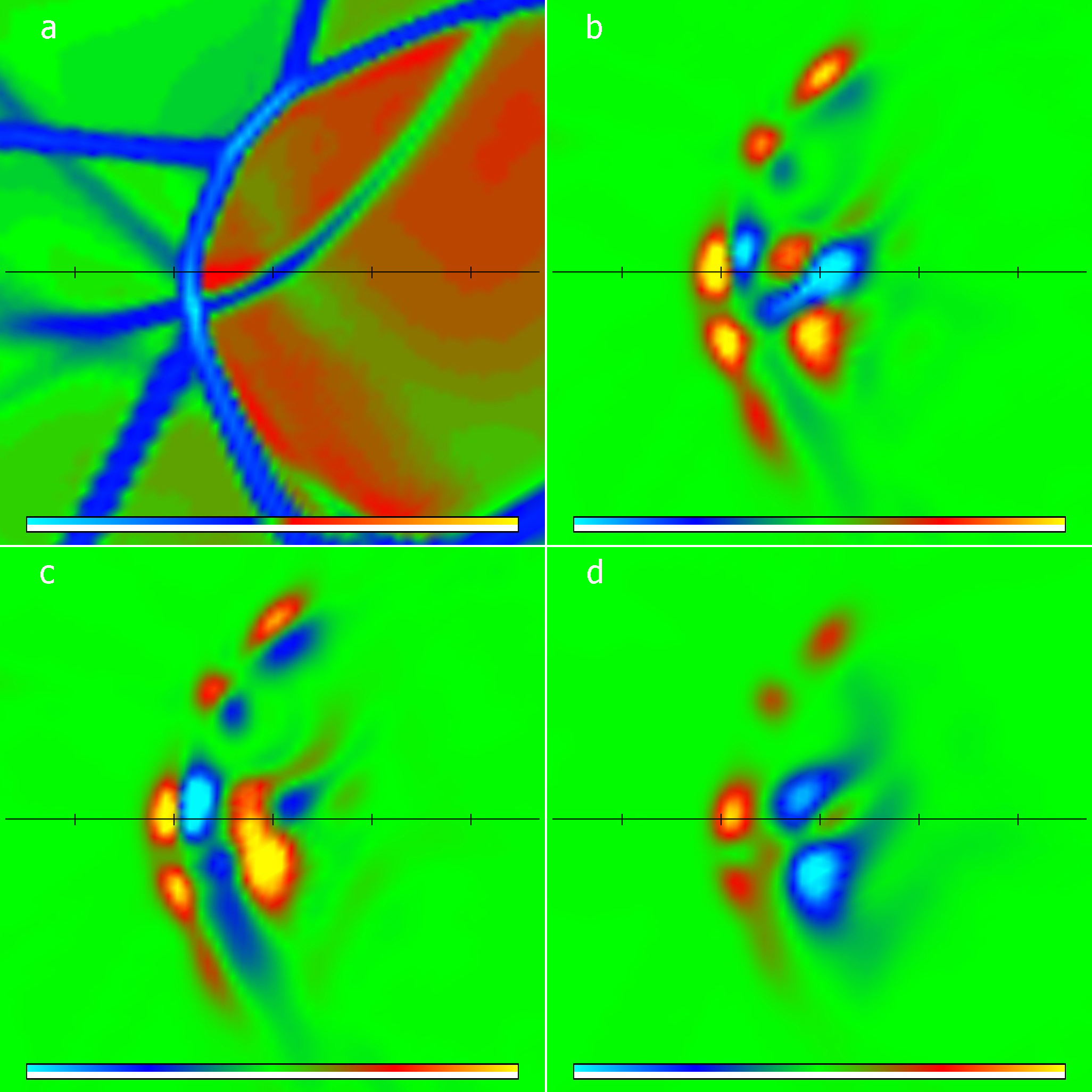}
   \caption{2D slice isolating intersecting shocks and enstrophy generation at $t=8$ in a compressive forcing simulation. {\bf  a)} $\nabla\cdot\vec u$, delineating the  pattern of shocks. {\bf b)} Total enstrophy source term, $\tilde F_{tot}$, of Favre-filtered flow as defined in Appendix \ref{filter}.  {\bf c)} Reynolds-stress enstrophy source term, $\tilde F_{sfs}$, of Favre-filtered flow. {\bf d)} Compressive enstrophy source term, $\tilde F_{comp}$. Flows are complex, but generally speaking ``upstream'' is to the left. Warm colors represent positive values; cool colors represent negative values, while green is zero. The horizontal line  with tick marks (separated by 0.5 length units) corresponds to the 1D cut in Figure \ref{scut}.}
    \label{compfig}
\end{figure}

\clearpage

\begin{figure}
    \centering
      \includegraphics[width=0.48\textwidth]{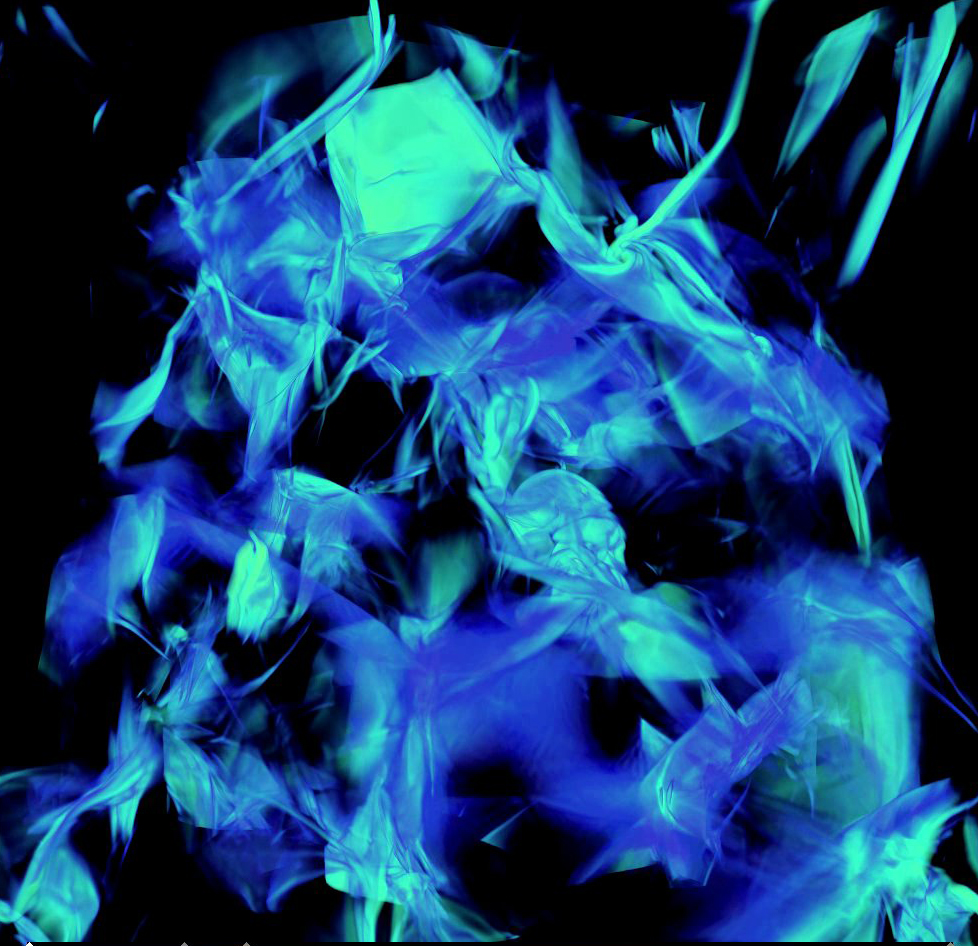}
      \includegraphics[width=0.48\textwidth]{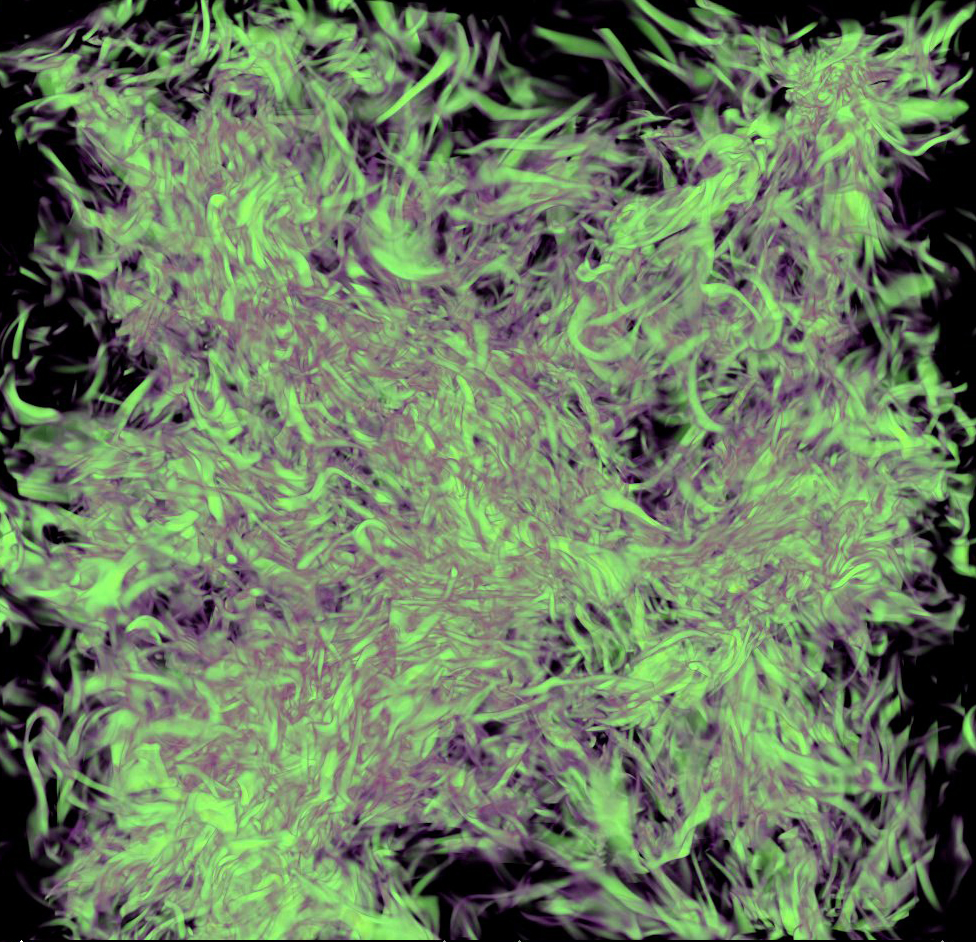}
    \caption{Magnetic energy density, $E_B$,  volume renderings in the C1K simulation at $t  =20$  (Left) and $t = 120$ (Right). ``Cool'' is weak; ``warm'' is strong. Opacities are chosen to isolate stronger field regions.}
    \label{compimages}
\end{figure}

\clearpage

   \begin{figure}
   \centering
   \includegraphics[width = 0.75\textwidth]{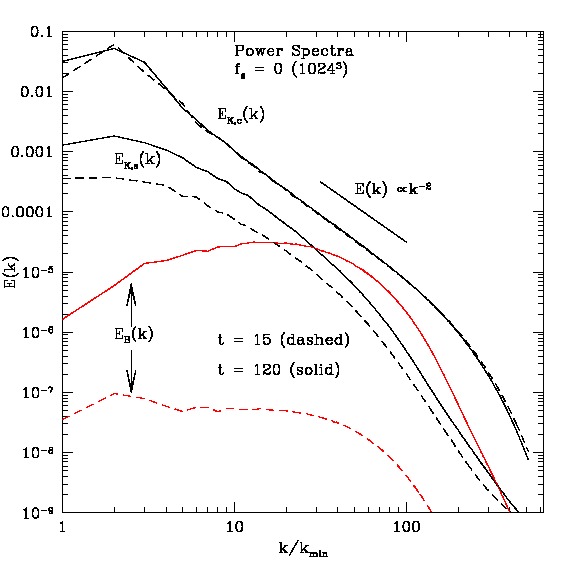}
   \caption{Power spectra of kinetic energy, $E_{K,c}(k)$ and $E_{K,s}(k)$, and magnetic energy, $E_B(k)$,  in the C1K simulation at $t = 15$ and $t = 120$.}
              \label{comp:spec}%
    \end{figure}

\clearpage

   \begin{figure}
   \centering
   \includegraphics[width = 0.75\textwidth]{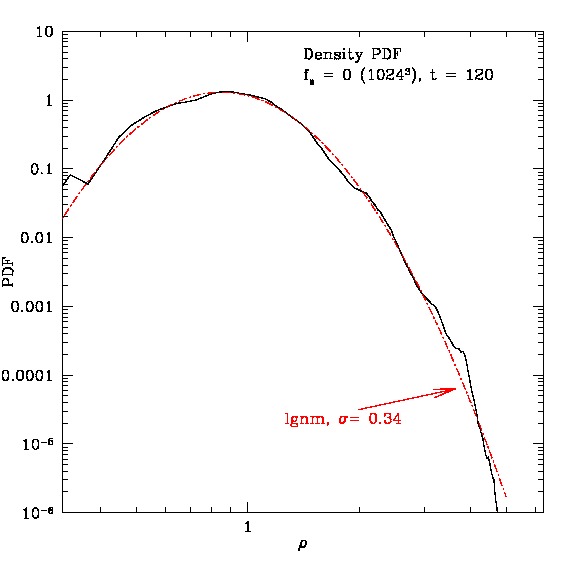}
   \caption{PDF of density in the C1K simulation at $t = 120$. Red dot-dash curve represents a  lognormal distribution fit.}
              \label{compdf}%
    \end{figure}

\clearpage

 \begin{figure}
   \centering
   \includegraphics[width = 0.8\textwidth]{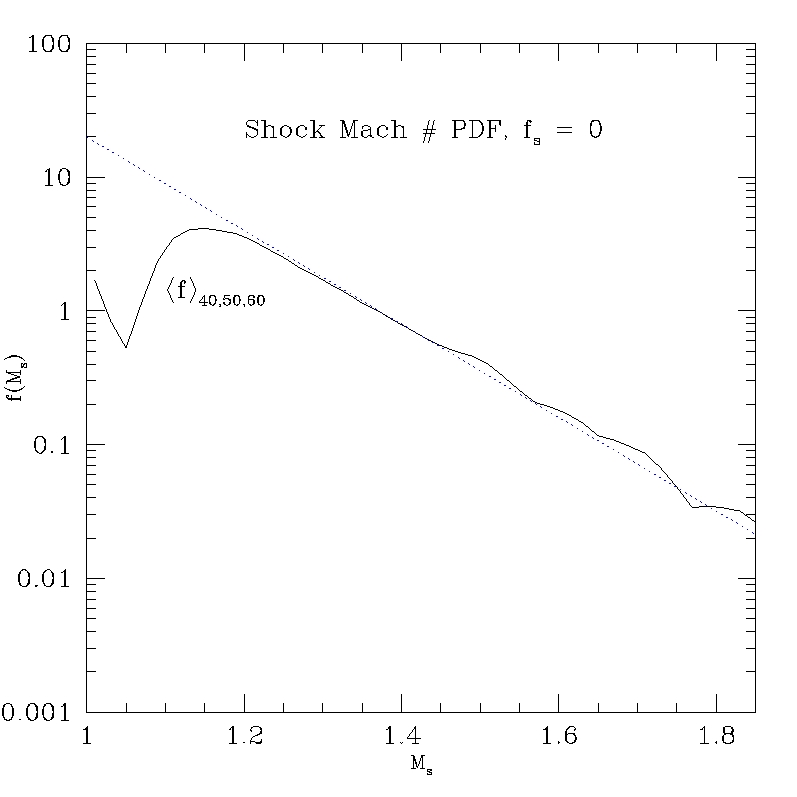}
   \caption{PDF of shock Mach numbers in the C1K simulation. The data curve  corresponds to an average for times $t = 40,50,60$. The dotted line represents an exponential PDF in the variable $\mathcal{M}_s-1$ with a characteristic $\mathcal{M}_{sc} =  1.125$.}
              \label{compmachdist}%
    \end{figure}

\clearpage

 \begin{figure}
   \centering
   \includegraphics[width = 0.8\textwidth]{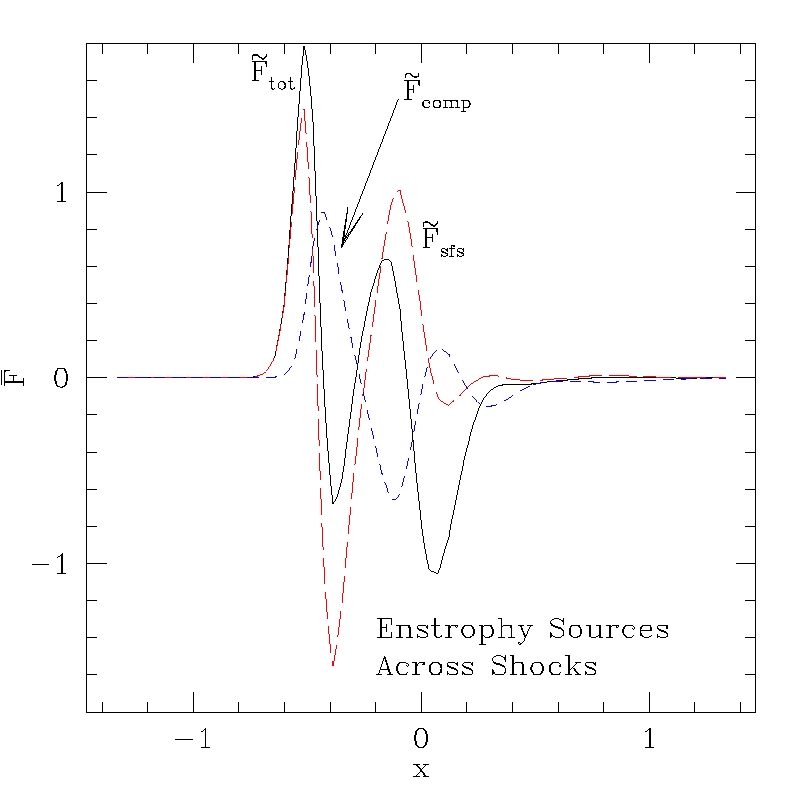}
   \caption{Plots of the total and dominant shock-concentrated enstrophy source terms from the filtered-flow enstrophy equation (\ref{enstrophy_source_favre}) along a horizontal cut through the shock complex shown in the 2D slice in Figure \ref{compfig}.}
              \label{scut}%
    \end{figure}


\end{document}